\pdfoutput=1
\newif\ifcomment
\newif\ifarxiv
\arxivtrue
\newif\ifplainart
\ifplainart
\documentclass[11pt]{article}
\usepackage{authblk}
\usepackage[nochapters]{classicthesis} 
\usepackage[comma,square,numbers,sort&compress]{natbib}
\else
\documentclass[%
 preprint,
 amsmath,
 amssymb,
 aps,
 pra,
]{revtex4-1}
\fi
\usepackage{dcolumn}
\usepackage{bm}
\usepackage{color}
\usepackage{graphicx}
\usepackage{amsmath}
\usepackage{amsfonts}
\usepackage{amssymb}
\usepackage[T1]{fontenc}
\usepackage{mathrsfs}
\usepackage{calrsfs}
\usepackage{pslatex}
\usepackage[allcolors=red]{hyperref}
\usepackage{url}
\usepackage{lineno}
\usepackage[utf8]{inputenc}
\usepackage[percent]{overpic}
\usepackage[separate-uncertainty=true,table-align-uncertainty=true]{siunitx}
\usepackage{array,multirow,graphicx}
\usepackage{booktabs}
\newcommand{\pp}           {pp}

\renewcommand{\AA}         {AA}

\newcommand{\NN}           {NN}
\newcommand{\pPb}          {pPb}

\newcommand{\pO}           {pO}

\newcommand{\PbPb}         {PbPb}

\newcommand{\AuAu}         {AuAu}
\newcommand{\XeXe}         {XeXe}
\newcommand{\NeNe}         {NeNe}
\newcommand{\OO}           {OO}

\newcommand{\pt}           {\ensuremath{p_{\rm T}}}

\newcommand{\TNN}          {\ensuremath{T_{\rm NN}}}

\newcommand{\rms}          {rms}
\newcommand{\eg}           {e.g.}
\newcommand{\ie}           {i.e.}

\newcommand{\snn}          {\ensuremath{\sqrt{s_{\rm NN}}}}

\newcommand{\sigmaNN}      {\ensuremath{\sigma_{\rm NN}}}
\newcommand{\sigmapPb}     {\ensuremath{\sigma_{\rm pPb}}}
\newcommand{\sigmapO}      {\ensuremath{\sigma_{\rm pO}}}
\newcommand{\sigmaPbPb}    {\ensuremath{\sigma_{\rm PbPb}}}
\newcommand{\sigmaOO}      {\ensuremath{\sigma_{\rm OO}}}
\newcommand{\sigmaNeNe}    {\ensuremath{\sigma_{\rm NeNe}}}

\newcommand{\dmin}         {\ensuremath{d_{\rm min}}}

\newcommand{\Ncoll}        {\ensuremath{N_{\rm coll}}}
\newcommand{\Npart}        {\ensuremath{N_{\rm part}}}

\newcommand{\Nmpi}         {\ensuremath{N_{\rm mpi}}}
\newcommand{\Nmpinn}       {\ensuremath{N^{\rm mpi}_{\rm NN}}}
\newcommand{\bnn}          {\ensuremath{b_{\rm NN}}}
\newcommand{\avg}[1]       {\ensuremath{\left\langle#1\right\rangle}}

\newcommand{\hrefurl}[1]   {\href{#1}{\url{#1}}}
\newcommand{\Refe}[1]      {Ref.~\cite{#1}}

\newcommand{\Tab}[1]       {Tab.~\ref{#1}}
\newcommand{\Fig}[1]       {Fig.~\ref{#1}}
\newcommand{\eFig}[1]      {extra Fig.~\ref{#1}}

\newcommand{\Eq}[1]        {Eq.~(\ref{#1})}
\newcommand{\Sec}[1]       {Sec.~\ref{#1}}
\newcommand{\App}[1]       {App.~\ref{#1}}
\newcommand{\Figure}[1]    {Figure~\ref{#1}}

\newcommand{\Appendix}[1]  {Appendix~\ref{#1}}
\newcommand{\Section}[1]   {Section~\ref{#1}}

\newcommand{\MCG}          {MCG}

\newcommand{\com}[1]       {}

\newcommand{\version}     {\href{http://www.hepforge.org/downloads/tglaubermc}{v3.3}}
\def\myTitle{Glauber predictions for oxygen and neon collisions\\ at energies available at the LHC}

\begin{document}
\ifplainart
\title{\myTitle}
\author{Constantin Loizides}  
\affil{\small CERN, 1211 Geneva 23, Switzerland}
\affil{\small Rice University, Houston, USA}
\date{\small February 3, 2026} 
\else
\title{\myTitle}
\author{Constantin Loizides}
\affiliation{CERN, 1211 Geneva 23, Switzerland}
\affiliation{Rice University, Houston, USA}
\date{\today}
\fi
\ifplainart
\maketitle
\fi
\begin{abstract}
The Glauber model is a widely used framework for describing the initial conditions in high-energy nuclear collisions. 
TGlauberMC is a Monte Carlo implementation of this model that enables detailed, event-by-event calculations across various collision systems. 
In this work, I present an updated version of TGlauberMC~(\version), which incorporates recent theoretical developments and improved parameterizations, especially relevant for small collision systems.
I focus on the oxygen--oxygen (\OO), neon--neon (\NeNe), and proton-oxygen (\pO) collisions at the Large Hadron Collider (LHC) in July 2025, where precise modeling of nuclear geometry and fluctuations is essential. 
The updated version includes revised nuclear density profiles and an enhanced treatment of nucleon substructure.
Geometrical cross sections for all relevant collision systems are calculated and initial-state observables are explored to provide predictions for particle production trends at $\snn=5.36$~TeV. 
In particular, a prediction for the centrality dependence of mid-rapidity multiplicity in \OO\ and \NeNe\ collisions is obtained. 
The updated code is publicly available to support the heavy-ion community with a robust and flexible tool for studying strongly interacting matter in small and intermediate-sized nuclear systems.

\ifarxiv
\vspace{0.15cm}
{\noindent\bf Revisions \&\& changes of the arXiv document and code:}\\
{\bf v1}, 08 Jul 2025: initial document, code v3.3 \\
{\bf v2}, 03 Sep 2025: submitted version, code v3.3.1 includes fix for Opar2 and new NNLOsat profile (``Osat'') \\
{\bf v3}, 03 Feb 2026: published version, code v3.3.2 includes smearing with TRENTO, and static functions
\fi
\end{abstract}
\ifplainart
\else
\maketitle
\fi
\ifarxiv
\tableofcontents
\newpage
\fi
\section{Introduction}
\label{sec:intro}
The study of heavy-ion collisions at ultra-relativistic energies has been instrumental in exploring the properties of the quark-gluon plasma (QGP), a state of matter believed to have existed shortly after the Big Bang~\cite{Busza:2018rrf}.
Experiments at the Relativistic Heavy Ion Collider~(RHIC) and the LHC have provided compelling evidence for the formation of QGP in large collision systems such as gold--gold~(\AuAu) and lead--lead~(\PbPb) at high collision energies~\cite{BRAHMS:2004adc,PHENIX:2004vcz,PHOBOS:2004zne,STAR:2005gfr,ALICE:2022wpn,CMS:2024krd}. 
Among the most prominent observables signaling QGP formation are collective flow phenomena, characterized by anisotropic flow coefficients such as $v_2$ and higher moments~\cite{Heinz:2013th}, and the phenomenon of jet quenching, in which high-energy partons lose energy as they traverse the dense medium~\cite{Cunqueiro:2021wls}.
Surprisingly, several QGP-like features have also been observed in small collision systems, such as proton-lead (\pPb) and even high-multiplicity proton-proton (pp) collisions~\cite{Loizides:2016tew,Nagle:2018nvi,Grosse-Oetringhaus:2024bwr}. 
These include long-range correlations and non-zero elliptic flow coefficients, raising questions about the minimal conditions required for QGP formation.
However, one signature remains notably absent in small systems: jet quenching. 
Despite observing various collective phenomena, experiments have not identified energy loss of high-$p_T$ partons in \pPb\ collisions~\cite{ALICE:2017svf,ATLAS:2022iyq}.
To further elucidate the system-size dependence of QGP signatures and to search for the possible onset of jet quenching in intermediate-sized systems, oxygen-oxygen (\OO) and neon-neon (\NeNe) collisions at a center-of-mass energy~($\snn$) of 5.36~TeV per nucleon pair provided by the LHC in July, 2025 are of great interest~\cite{Citron:2018lsq,Brewer:2021kiv}. 
These collision systems provide a crucial bridge between small and large systems, allowing precise control measurements of observables such as jet quenching and elliptic flow~\cite{Lim:2018huo,Sievert:2019zjr,Huang:2019tgz,Huss:2020dwe,Jia:2022ozr}. 
Specifically, the role of $\alpha$ clusters in small nuclei including $^{16}$O has long been discussed and may be imprinted in the final-state flow measurements~\cite{Rybczynski:2017nrx,Rybczynski:2019adt,Li:2020vrg,Summerfield:2021oex,Ding:2023ibq,YuanyuanWang:2024sgp,Zhang:2024vkh}.
Recently, comparing $v_2$ in \NeNe\ to $v_2$ in \OO\ collisions was highlighted to provide a precision test of the role of initial geometry and the QGP paradigm~\cite{Giacalone:2024luz}.
Additionally, proton--oxygen (\pO) collisions at a $\snn=9.62$~TeV were provided, primarily to improve hadronic interaction models used in the interpretation of extensive air showers in cosmic-ray physics\cite{Dembinski:2020dam}, but they also present a unique opportunity to investigate QGP-related phenomena.


Interpretation of high-energy nucleus collision data typically depends on the use of classical Glauber Monte Carlo~(MCG) models to describe the initial condition~\cite{Miller:2007ri,dEnterria:2020dwq}.
The general approach is to deduce properties of QCD matter by  measuring the distributions of a variety of observables in \AA\ collisions, and comparing them to the same measurements in \pp\ collisions, where a QGP is not expected to be produced. 
To compare colliding systems of different sizes requires hence appropriate normalization of the measured distributions by using \eg\ the number of participating nucleons~($\Npart$), the number of independent binary nucleon-nucleon~(\NN) collisions~($\Ncoll$), the medium transverse area ($A_\perp$), or the overlap eccentricity~$\epsilon_n$ (given by the $n^{\rm th}$ moments of its azimuthal spatial distribution).

Widely used and publicly available \MCG\ models are the TGlauberMC~\cite{Loizides:2014vua,Loizides:2016djv,Loizides:2017ack}, which developed from the original PHOBOS implementation~\cite{Alver:2008aq} and GLISSANDO codes~\cite{Broniowski:2007nz,Rybczynski:2013yba,Bozek:2019wyr}.
The Glauber initial condition is widely used in simulations of heavy-ion collisions as provided by TRENTO~\cite{Moreland:2014oya}, Trajectum~\cite{Nijs:2021clz} and MUSIC~\cite{Paquet:2015lta}, as well as by event generators such as HIJING~\cite{Wang:1991hta}, AMPT~\cite{Lin:2004en}, URQMD~\cite{Bass:1998ca} and EPOS~\cite{Werner:2010aa}.
An alternative description of the initial condition is provided by IP-Glasma~\cite{Schenke:2018fci}.

In this article, an updated version~(\version) of the TGlauberMC model is used to predict properties of \OO, \NeNe, as well as \pPb\ and \PbPb\ collisions at 5.36~TeV, and \pO\ collisions at 9.62~TeV.
In particular, I focus on the calculation of the expected cross sections and discuss aforementioned associated initial-state quantities, such as $\Npart$, $\Ncoll$ and eccentricity distributions for \OO\ and \NeNe\ collisions including the ratio of eccentricities for \NeNe\ to \OO\ as a proxy for the respective ratio of $v_{2}$ in these two systems. 
Finally, particle production at 5.36~TeV is discussed by outlining some features derived from the multiple parton interaction~(MPI) picture~\cite{Sjostrand:2017cdm}, and use existing data at 5.36 TeV to predict the multiplicity in \OO\ and \NeNe\ collisions.
As for previous versions of the model, the source code for TGlauberMC~(\version) has been made publicly available at HepForge~\cite{glaucode}.

The remainder of the article is divided in the following sections.
\Section{sec:MCG} briefly introduces the general \MCG\ approach.
\Section{sec:transoverlap} focuses on the treatment of the transverse nucleon--nucleon overlap.
\Section{sec:nuclearprof} discusses the available parameterizations and calculations of $^{16}$O and $^{20}$Ne nuclear densities.
\Section{sec:result} presents the predictions for the hadronic cross sections at 5.36 and 9.62~TeV.
\Section{sec:init} discusses the $\Npart$, $\Ncoll$ and eccentricity distributions for \OO\ and \NeNe.
\Section{sec:partprod} focuses on particle production using the MPI model..
\Section{sec:summary} provides a concise summary of the main findings of this article. 
\Appendix{app:guide} describes the improvements of TGlauberMC~(\version) and provides a short user's guide. 
\Appendix{sec:trnucgen} introduces a new software library called ``Trajectum Nucleus Generator (TrNucGen)'' to provide nuclear densities calculated by Trajectum.
\ifarxiv
\Appendix{sec:addfigs} provides a set of extra figures not available in the published version.
\fi

\section{The Monte Carlo Glauber approach}
\label{sec:MCG}
Monte Carlo Glauber~(\MCG) calculations evaluate the cross section of two nuclei with $A$ and $B$ nucleons stochastically by distributing them in coordinate space according to the corresponding nuclear densities, separated by an impact parameter $b$ sampled from $\mathrm{d}\sigma/\mathrm{d}b \propto b$.
For simplicity, the impact-parameter variable $b$ or $\bnn$ denotes the distance between nuclei~(or nucleons) in the transverse plane.
The nuclear transverse profiles of the nuclei are usually approximated with parametrizations of charge density distributions extracted from low-energy electron-nucleus scattering experiments~\cite{DeVries:1987atn}.
For large nuclei, they are usually described by a 3-parameter Fermi~(3pF) distribution 
\begin{equation}
\rho(r)=\rho_0\left(1+w\frac{r^2}{R^2}\right)/\left(1+\exp(\frac{r-R}{a})\right)
\label{eq:3pF}
\end{equation}
with half-density radius~$R$ and diffusivity~$a$, and $w\neq0$ describing small deviations from a spherical shape.
Following the eikonal approximation~\cite{Capel:2019zor}, the collision is treated as a sequence of independent binary nucleon-nucleon collisions, \ie\ the nucleons travel on straight-line trajectories and their interaction probability does not depend on the number of collisions they have suffered before. 
In the so-called ``hard-sphere''~(HS) approximation, an interaction takes place between two nucleons if the transverse distance $d$ (or generally the transverse nucleon--nucleon impact parameter $\bnn$) between their centers satisfies $d < \sqrt{\sigmaNN/\pi}$.

In the \MCG\ approach, the number of participants and collisions are computed by counting the number of nucleons with at least one collision, and the total number of nucleon--nucleon collisions, respectively. 
The collision cross section is the fraction of simulated collisions with $\Ncoll>0$ of all simulated collisions.
Quantities such as the eccentricity~\cite{PHOBOS:2006dbo}, the triangularity~\cite{Alver:2010gr}, and higher moments~\cite{Teaney:2010vd} of the collision region at impact parameter $b$ are given by
\begin{equation}
\varepsilon_{n}(b)=\frac{\left<r^n\cos(n\phi-n\psi)\right>}{\left<r^n\right>}
\label{eq:ecc}
\end{equation}
where $n$ denotes the moment, $r=\sqrt{x^2+y^2}$ and $\psi=\tan^{-1}\frac{y}{x}$.
The averages are computed by considering the central positions of either, participant nucleons or binary nucleon--nucleon collisions, or of an admixture of the two, and additionally may be smeared to model matter production~\cite{PHOBOS:2007vdf}.

\begin{figure}[ht!] 
\includegraphics[width=8cm]{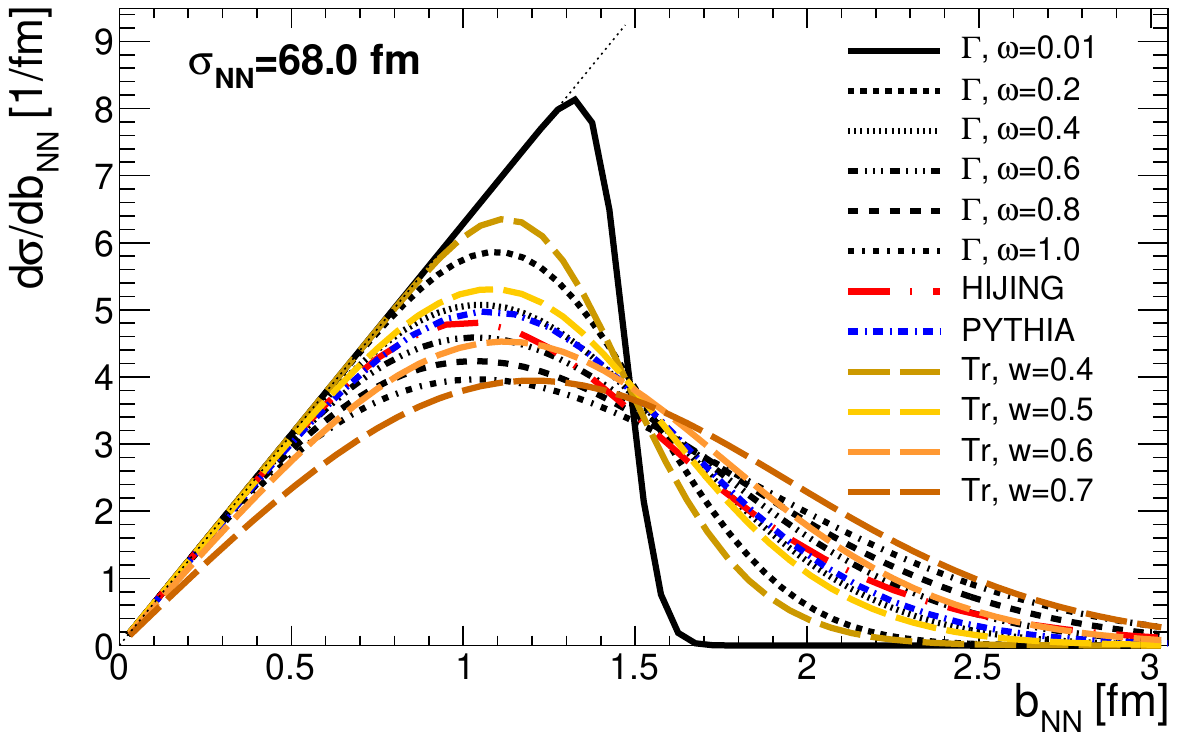}
\includegraphics[width=8cm]{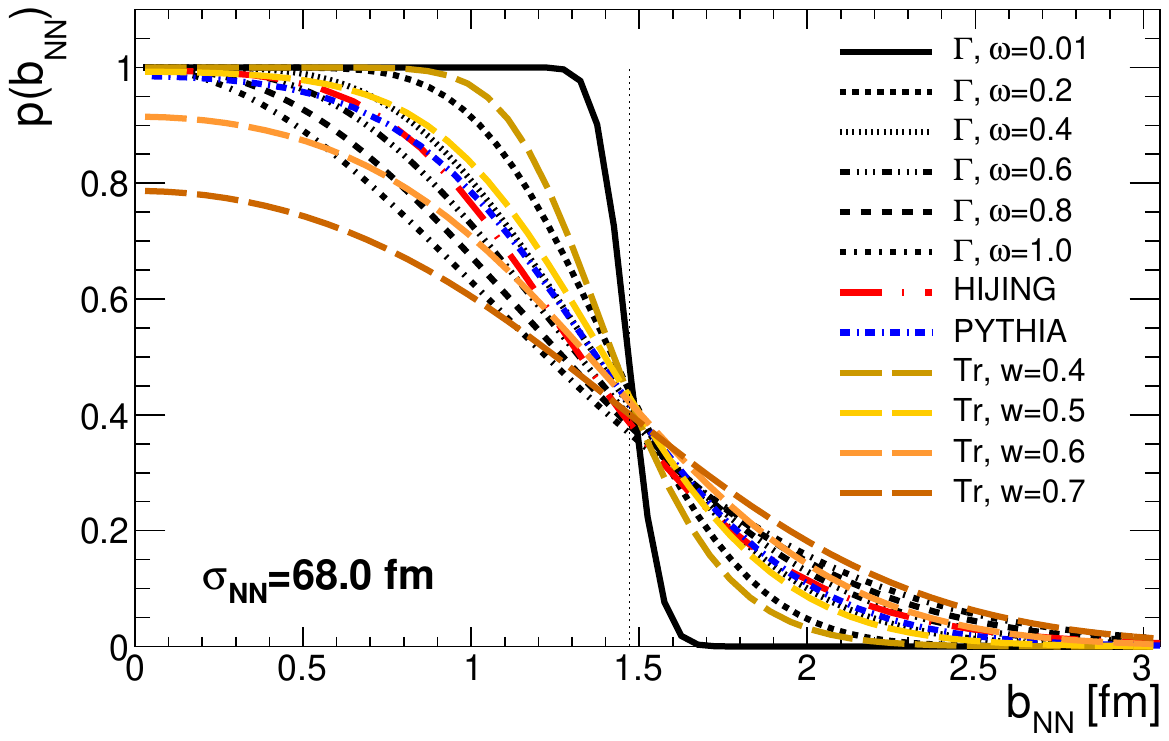}
\caption{
Nucleon--nucleon impact parameter distribution~(left) and interaction probability~(right) at 5.36~TeV shown for the different transverse overlap functions ($\Gamma$-parameterization with $0\le\omega\le1$, TRENTO with $0.4\le w\le0.7$, HIJING and PYTHIA tune Monash) as discussed in the text. The thin dashed line indicates the hard-sphere approximation.
}
\label{fig:oproftglaubermc}
\end{figure}

\section{Transverse overlap function}
\label{sec:transoverlap}
Several extensions to the hard-sphere approximation have been explored to account for the fact that the nucleon matter profile is not a black disc~\cite{dEnterria:2010xip}.
A commonly applied extension, referred to as TRENTO~\cite{Moreland:2014oya},
is to assume the nuclear matter profile of a nucleon to be a Gaussian with a certain width $w$, so that the nuclear overlap function
\begin{equation}
\TNN(\bnn) \propto \exp \left[-\left(\bnn/B\right)^{m} \right] 
\end{equation}
is also Gaussian, i.e.\ $m=2$ and $B=2w$.
For some settings~(``bProfile = 3''~\cite{Sjostrand:2017cdm}) PYTHIA assumes $B=1$~fm as characteristic size of the proton, and uses $m$ to interpolate between Exponential~($m\to1$) and Gaussian~($m\to 2$) shape. 
The widely-used ``Monash'' tune sets $m=1.85$~\cite{Skands:2014pea}.
HIJING~\cite{Wang:1991hta} uses an overlap function based on modified Bessel functions of the 3rd kind, 
\begin{equation}
\TNN({\bnn}) \propto (\mu\,\bnn)^3\,K_3(\mu\,\bnn) \,,
\label{eq:overlaphijing}
\end{equation} 
with $\mu={1.5}/{\sqrt{\sigmaNN}}$.
In the Eikonal formalism, the probability to interact is then given as $p(\bnn)=1-\exp\left(-k\,\TNN(\bnn)\right)$ where $k$ is obtained by requiring $2\pi\,\int p(\bnn)\,\bnn\,{\rm d}\bnn=\sigmaNN$.
At LHC collision energies, a practical parametrization~\cite{Rybczynski:2013mla} based on the Euler $\Gamma(z)$ and incomplete $\Gamma(\alpha,z)$ functions, in the following also called $\Gamma$-parametrization, is
\begin{equation}
p(\bnn) \propto \Gamma\left(1/\omega,\frac{\bnn^2}{d^2\omega}\right)\bigg/\Gamma(1/\omega)\,,
\label{eq:NN_coll_profile}
\end{equation}
where $\omega$ interpolates between the hard-sphere ($\omega\to 0$) and Gaussian ($\omega\to 1$) cases.

The different choices for the nucleon--nucleon impact parameter distributions and corresponding interaction probabilities are shown in \Fig{fig:oproftglaubermc} for the $\Gamma$-parameterization with $0\le\omega\le1$, TRENTO with $0.4\le w\le0.7$, HIJING and PYTHIA tune Monash.
Beyond the hard-sphere approximation, there is a quite a large parameter space, leading to a significant uncertainty resulting from the dependence on the parameterization. 
However, HIJING, PYTHIA, as well as the $\Gamma$-parameterizations for $\omega\sim0.5$ and TRENTO for $w\sim0.5$ lead to relatively similar distributions, quite distinct from the hard-sphere approximation.

\begin{figure}[th!] 
\includegraphics[width=10cm]{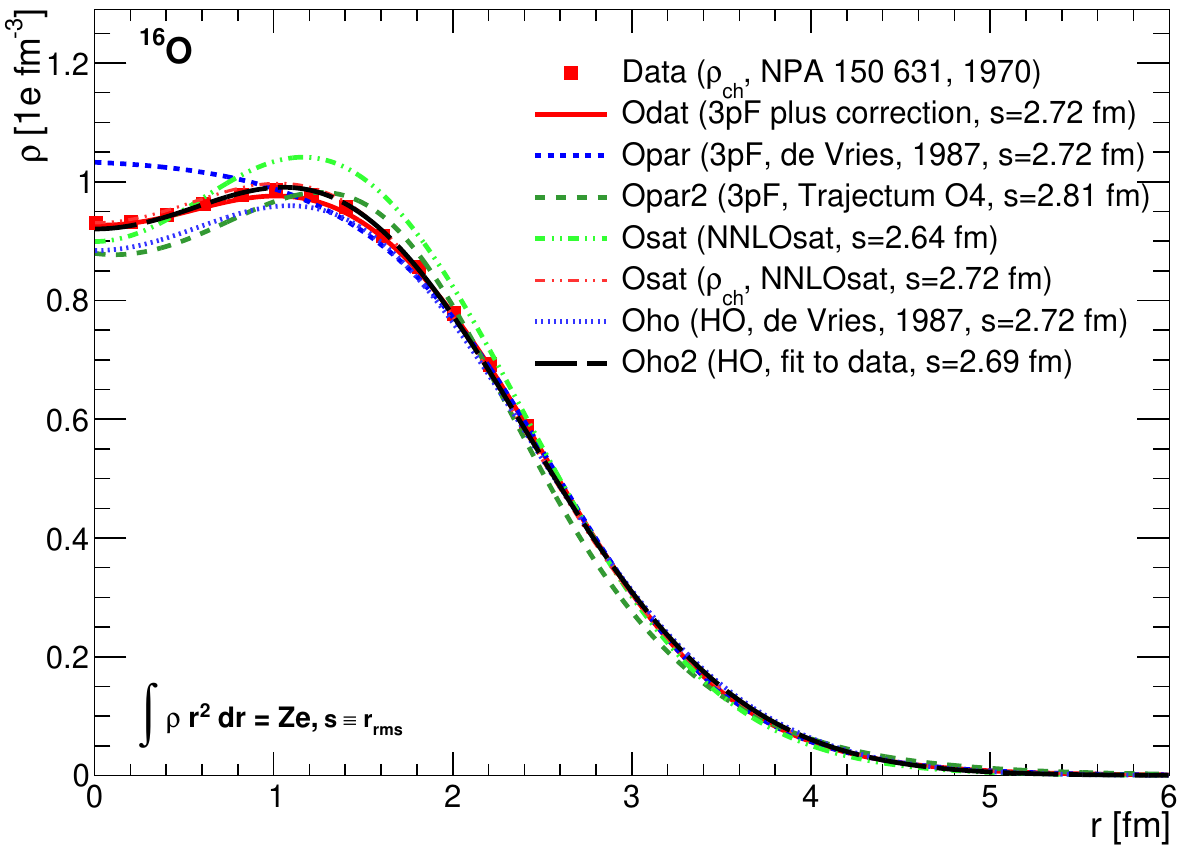}
\caption{Nuclear charge density for $^{16}$O versus radius for various descriptions compared to data from electron scattering~\cite{Sick:1970ma}. As explained in the text, ``Oho2'' is a parameterization based on the HO model, whose
parameters are fit to the data, while ``Odat'' was already provided in \cite{Sick:1970ma}. 
``Osat'' is the exact result of the NNLOsat calculation~\cite{Soma:2019bso}), and ``Opar2'' the respective 3pF fit.
``Opar'' and ``Oho'' profiles were previously included in TGlauberMC. 
The distributions are normalized as $\int \rho\,r^2\,{\rm d}r=Ze$.
The respective $r_{\rm rms}$ values are given in the legend, denoted as $s$.
}
\label{fig:oprofdata}
\end{figure}

\section{Nuclear density profiles for oxygen and neon}
\label{sec:nuclearprof}
The standard method to construct the nucleon configurations for \MCG\ calculations is to sample the nucleon positions from $r^2\rho$, where the nuclear density $\rho$ may include angular deformations~\cite{Loizides:2014vua}.
To mimic a hard-core repulsion potential between nucleons, a minimum inter-nucleon separation~($\dmin$) of $0.4$~fm between their centers is enforced when sampling the positions of the nucleons inside a nucleus.
In order to ensure that the center-of-mass of each constructed nucleus is at $(0,0,0)$, the nucleons are individually ``shifted'' by the necessary amount, after all nucleon positions were sampled.
Both, imposing the hard-core repulsion as well as the re-centering, induce a bias from the original $\rho$ distribution.
As discussed in \Refe{Loizides:2017ack}, this can be overcome by a ``reweighting'' procedure, modifying the original $\rho$ distribution to counteract the bias.
However, for $\dmin=0.4$~fm, and recentering via rotation around the $z$-axis~(recentering option 4 in TGlauberMC), residual differences are negligible. 

\begin{table}[hbt!]
\centering
\caption{Nuclear parameters for oxygen and neon parameterizations provided by TGlauberMC~(\version) with given names as in the code.
The 3pF parameterization is given in \Eq{eq:3pF}.
The HO parameterization is given in \Eq{eq:ho}.
The last two configurations used deformed profiles, with deformation values as given in the comment field, and defined in Trajectum~\cite{Giacalone:2024luz}.
The prefix ``TR'' indicates the profile is taken from Trajectum, where the name following the prefix denotes the system and version number as used in Trajectum.
Included as ``Osat'' is also the exact profile from the ${\rm NNLO}_{\rm sat}$ chiral Hamiltonian calculation~\cite{Soma:2019bso}. 
For completeness, ``O'' is also listed which uses explicit nucleon configurations~\cite{Lonardoni:2017hgs} as introduced in~\cite{Lim:2018huo}.
}
\begin{tabular}{|l|c|c|c|c|c|l|}
\hline
\textbf{Name} & $A$ & $Z$ & $R$ [fm] or $\alpha$ & $a$ [fm] & $w$ & \textbf{Comments} \\
\hline
Odat    & 16 & 8  & 2.608  & 0.513  & -0.051 & Original parametrization~(3pF plus corrections) from \cite{Sick:1970ma} \\
Opar    & 16 & 8  & 2.608  & 0.513  & -0.051 & 3pF parameterization~\cite{DeVries:1987atn} (same as TR\_OXYGEN\_V3) \\
Opar2   & 16 & 8  & 1.850  & 0.497  & 0.912  & 3pF parameterization (TR\_OXYGEN\_V4, fit to ${\rm NNLO}_{\rm sat}$~\cite{Soma:2019bso}) \\
Oho     & 16 & 8  & 1.544  & 1.833  & -      & Harmonic Oscillator~\cite{DeVries:1987atn} \\
Oho2    & 16 & 8  & 1.506  & 1.819  & -      & Harmonic Oscillator (new fit to data from \cite{Sick:1970ma}) \\\hline
Osat    & 16 & 8  & -      & -      & -      & Exact NNLO$_{\rm sat}$ profile from \cite{Soma:2019bso}, use instead of Opar2 \\\hline
O       & 16 & 8  & -      & -      & -      & Nucleon configurations~\cite{Lim:2018huo}~(same as in TGlauberMC v2.7) \\\hline
Ne      & 20 & 10 & 2.805  & 0.571  & -      & 2pF parametrization~\cite{DeVries:1987atn} \\
Ne2     & 20 & 10 & 2.740  & 0.572  & -      & 2pF parametrization~\cite{DeVries:1987atn} \\
Ne3     & 20 & 10 & 2.791  & 0.698  & -0.168 & 3pF parametrization~\cite{DeVries:1987atn} \\
NeTr2   & 20 & 10 & 2.805  & 0.571  & -      & TR\_NEON\_V2, (same as Ne with $\beta_2 = 0.721$ from \cite{Pritychenko:2013gwa}) \\
NeTr3   & 20 & 10 & 2.724  & 0.498  & -      & TR\_NEON\_V3, $\beta_2 = 0.4899$, $\beta_3 = 0.2160$, $\beta_4 = 0.3055$ \\
\hline
\end{tabular}
\label{tab:gloxygen}
\end{table}

\Figure{fig:oprofdata} shows the nuclear charge density for $^{16}$O for data from electron scattering~\cite{Sick:1970ma}, including the original parameterization~(``Odat''), as well as 3pF distributions with various parameters available in the literature~(``Opar'', and ``Opar2'').
\Refe{DeVries:1987atn} also gives a parameterization based on the (modified) Harmonic Oscillator~(HO) model
\begin{equation}
\rho(r) = \rho_0 \left( 1 + \alpha \left(x/a\right)^2 \exp{-\left(x/a\right)^2}\right)
\label{eq:ho}
\end{equation}
labeled ``Oho'' in the figure.
For this work, I also provide a new HO fit to the data, called ``Oho2'', leading to $\alpha=1.506\pm0.023$ and $1.819\pm0.004$ with $\chi^2/n_{\rm dof}\approx1$ assuming about 1\% relative error on the data points.
While it overall has a worse $\chi^2$ than ``Odat'', it only has two parameters and can be used as alternative and allows systematic variation using the uncertainties provided on the fit parameters.
To reflect the charge density, the distributions shown in \Fig{fig:oprofdata} are normalized as $\int \rho\,r^2\,{\rm d}r=Ze$.
A detailed list of all parameterizations shown in \Fig{fig:oprofdata} is given in \Tab{tab:gloxygen}.
The $r_{\rm rms}=\sqrt{\langle r\rangle}$ values are reported in the legend of the figure, and differ by nearly up to 5\%.
The reason is that the data do not constrain the tail of the distribution, which was found to affect the $r_{\rm rms}$ by a few percent, and thus experimental results between $2.65\pm0.04$ and $2.73\pm0.03$ have been reported~\cite{Sick:1970ma}.
Furthermore small differences can also be expected from inclusion of neutrons in the density distributions, and the use of point-like protons ignoring the proton form factor.
However, the distribution ``Opar2'' which results from a fit to the ${\rm NNLO}_{\rm sat}$ calculation~\cite{Soma:2019bso} exhibits a significantly larger $r_{\rms}$ than the data (and than reported in~\cite{Soma:2019bso}), and hence is not further considered here.
Instead, the exact nuclear density obtained from the ${\rm NNLO}_{\rm sat}$ chiral Hamiltonian calculation~\cite{Soma:2019bso} is included and also shown in the figure. The nuclear charge density directly calculated by ${\rm NNLO}_{\rm sat}$ perfectly agrees with the data (also shown in the figure).

\begin{table}[h]
\centering
\caption{Oxygen-16 and Neon-20 NLEFT and PGCM calculations taken from Trajectum~\cite{Giacalone:2024luz}.}
\begin{tabular}{ll}
\hline
Profile & Description \\
\hline
TR\_OXYGEN\_V12 & NLEFT pinhole, $\pm$ weights, with periodicity ambiguities resolved \\
TR\_OXYGEN\_V14 & PGCM, constraints imposed on projected states, without explicit $\alpha$ clustering~($C=0$) \\
TR\_OXYGEN\_V15 & PGCM, constraints imposed on projected states, including explicit $\alpha$ clustering~($C=1$) \\
\hline 
TR\_NEON\_V11 & NLEFT pinhole, $\pm$ weights, with periodicity ambiguities resolved \\
TR\_NEON\_V13 & PGCM, constraints imposed on projected states, without explicit $\alpha$ clustering~($C=0$) \\
TR\_NEON\_V14 & PGCM, constraints imposed on projected states, with explicit $\alpha$ clustering~($C=1$) \\
\hline
\end{tabular}
\label{tab:troxygen}
\end{table}

Instead of sampling nucleon positions from average $\rho$ distributions, explicit calculations of the nucleon configuration based on the many-body wave function, is desired to account for inter-nucleon correlations, in particular important for small and deformed nuclei~\cite{Dobaczewski:2025rdi}.
In this case, an explicit $\dmin$ criterion between nucleons should not be imposed. 
Previous work~\cite{Lim:2018huo} used an ab-initio quantum Monte Carlo calculation of the Oxygen nucleon configurations with two- and three-body local chiral potentials at next-to-next-to-leading order~\cite{Lonardoni:2017hgs}, which were available in TGlauberMC~(since v2.7, named ``O'', see \Tab{tab:gloxygen}).
Recently, Oxygen and Neon configurations where derived within the framework of Nuclear Lattice Effective Field Theory~(NLEFT) and the ab-initio Projected Generator Coordinate Method~(PGCM), and used as input for the initial conditions in Trajectum~\cite{Giacalone:2024luz}.
The NLEFT framework is well suited to probe collective phenomena in the ground states of nuclei~\cite{Lu:2018bat}.
Similarly to \cite{Summerfield:2021oex}, the positions of the nucleons were traced using the pinhole~\cite{Elhatisari:2017eno} algorithm.
In this way, the NLEFT configurations, which can be obtained keeping track of positive and negative weights, and with resolving ambiguities introduced by the underlying lattice periodicity, capture short-range correlations without explicitly enforcing $\alpha$-clusters.
In the PGCM approach~\cite{Frosini:2021ddm,Huther:2019ont}, a set of mean-field states with different quadrupole and octupole deformations is first generated to explore a range of possible nuclear shapes. 
Each of these intrinsic configurations is then projected onto states with physical particle number and angular momentum to restore the symmetries broken at the mean-field level. 
A collective wavefunction is constructed by mixing the projected states through a variational procedure that minimizes the total energy of the system.
The final nucleon configurations are then constructed with or without explicitly enforcing $\alpha$-clustering, denoted as ``PGCM,~$C=1$'' and ``PGCM,~$C=0$'', respectively.
A summary of the physically meaningful nucleon configurations obtained by the NLEFT and PGCM methods is given in \Tab{tab:troxygen}.
For more details, see the extra material provided in~\Refe{Giacalone:2024luz}.

\begin{figure}[t] 
\includegraphics[width=8.0cm]{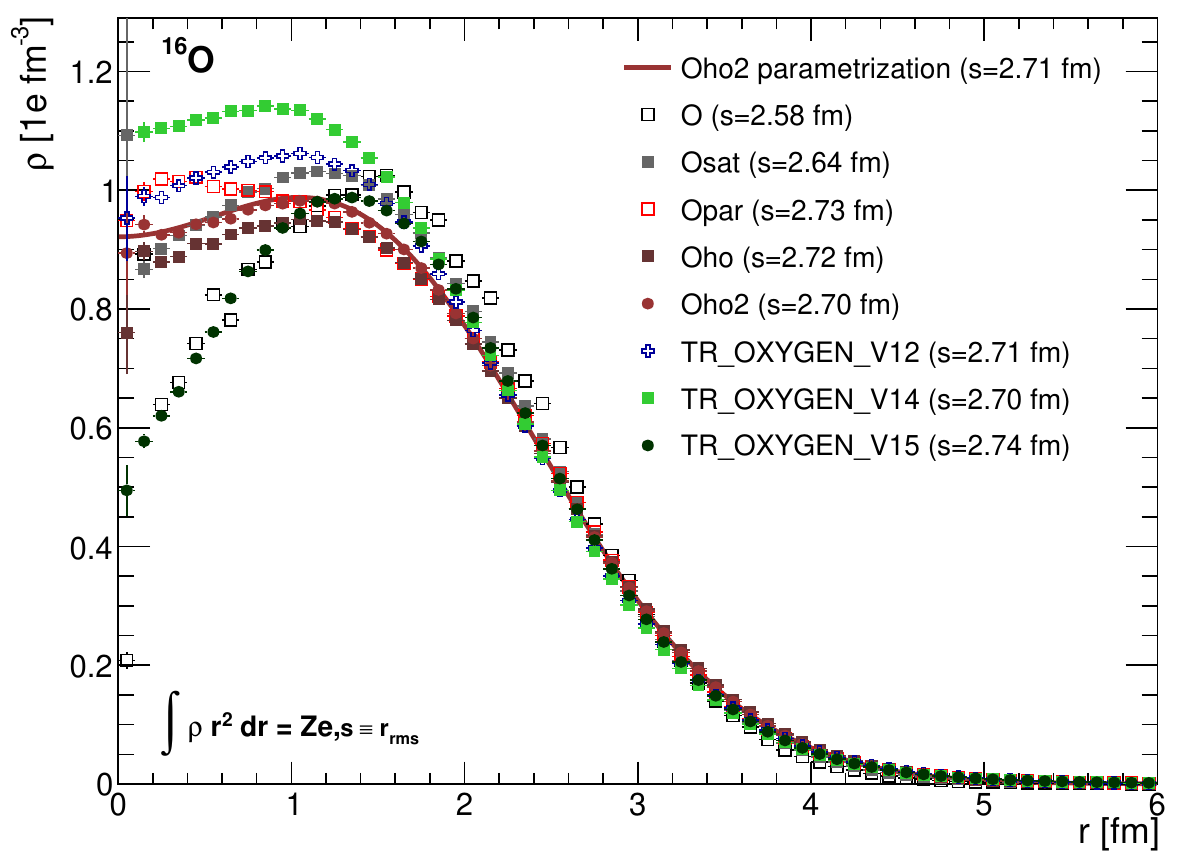} 
\includegraphics[width=8.0cm]{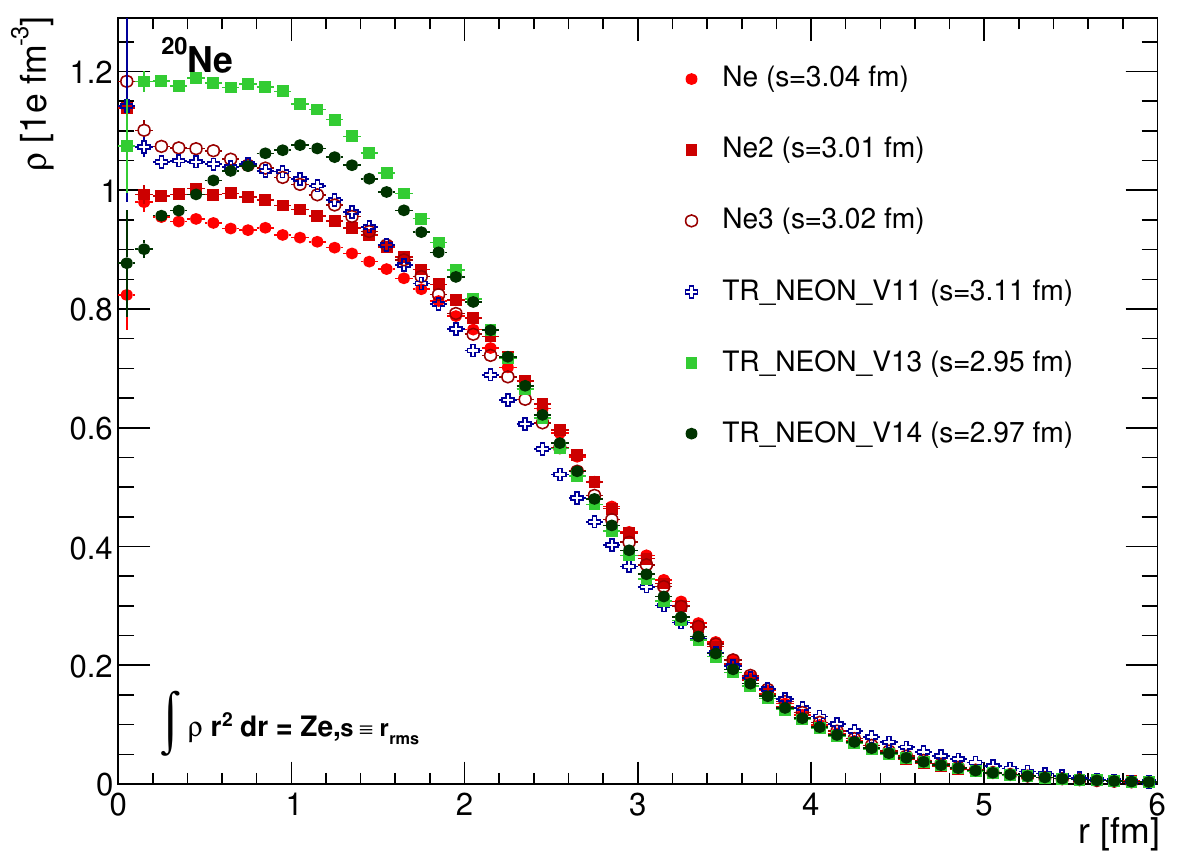}
\caption{Nuclear charge density for $^{16}$O (left) and $^{20}$Ne~(right) versus radius obtained with TGlauberMC for the various parameterizations discussed in the text. The respective $r_{\rm rms}$ values are given in the legend, denoted as $s$. In case of oxygen, the calculated densities are compared to the fit of the data~(``Oho2'').}
\label{fig:oprofdata2}
\end{figure}

The nuclear charge densities for $^{16}$O obtained with TGlauberMC using the nuclear charged densities listed in \Tab{tab:gloxygen} and the explicit nucleon configurations for NLEFT and PCGM listed in \Tab{tab:troxygen} in the left panel of \Fig{fig:oprofdata2}, compared to the parameterization ``Oho2'' introduced above.
As can be seen, there is a significant spread at lower radii between the different cases, which on the one hand reflect the different $\rho$ parameterizations shown in \Fig{fig:oprofdata}, and on the other hand reflects the differences between cases that implicitly or explicitly include $\alpha$-clusters~(``O'' and ``TR\_OXYGEN\_V15'') exhibiting a dip at small $r$.
In the right panel of \Fig{fig:oprofdata2}, for completeness, meaningful profiles for $^{20}$Neon are shown, similarly revealing a large spread, with again configurations with explicit $\alpha$-clustering~(``TR\_NEON\_V14'') exhibit a dip at small $r$ values.
Not shown but mentioned in \Tab{tab:gloxygen} are ``NeTr2'' and ``NeTr3'' because the former has a nearly 10\% larger $r_{\rm rms}$ than reported from the data and the latter was fit to an unprojected version of the PCGM calculation.

\section{Cross section results}
\label{sec:result}
To compute the expected \OO, \NeNe\ and \pO\ geometrical cross sections at 5.36 and 9.62~TeV, one needs the values for $\sigmaNN$ at the respective nucleon--nucleon center-of-mass beam energy as input to the \MCG\ calculation. 
The $\sigmaNN$ can be obtained from the parametrization, discussed in~\cite{dEnterria:2020dwq},
\begin{equation}
\sigmaNN(\sqrt{s}/{\rm GeV}) = a + b \,\ln^{n}(s/{\rm GeV}^2)\,,
\label{eq:signnfunc}
\end{equation}
with $a=28.84\pm0.52$, $b=0.0458\pm0.0167$ and $n=2.374\pm0.123$, resulting in $\sigmaNN=68.0\pm1.2$~mb at 5.36~TeV and $\sigmaNN=74.7\pm1.5$~mb at 9.62~TeV.

\begin{figure}[t] 
\includegraphics[width=8cm]{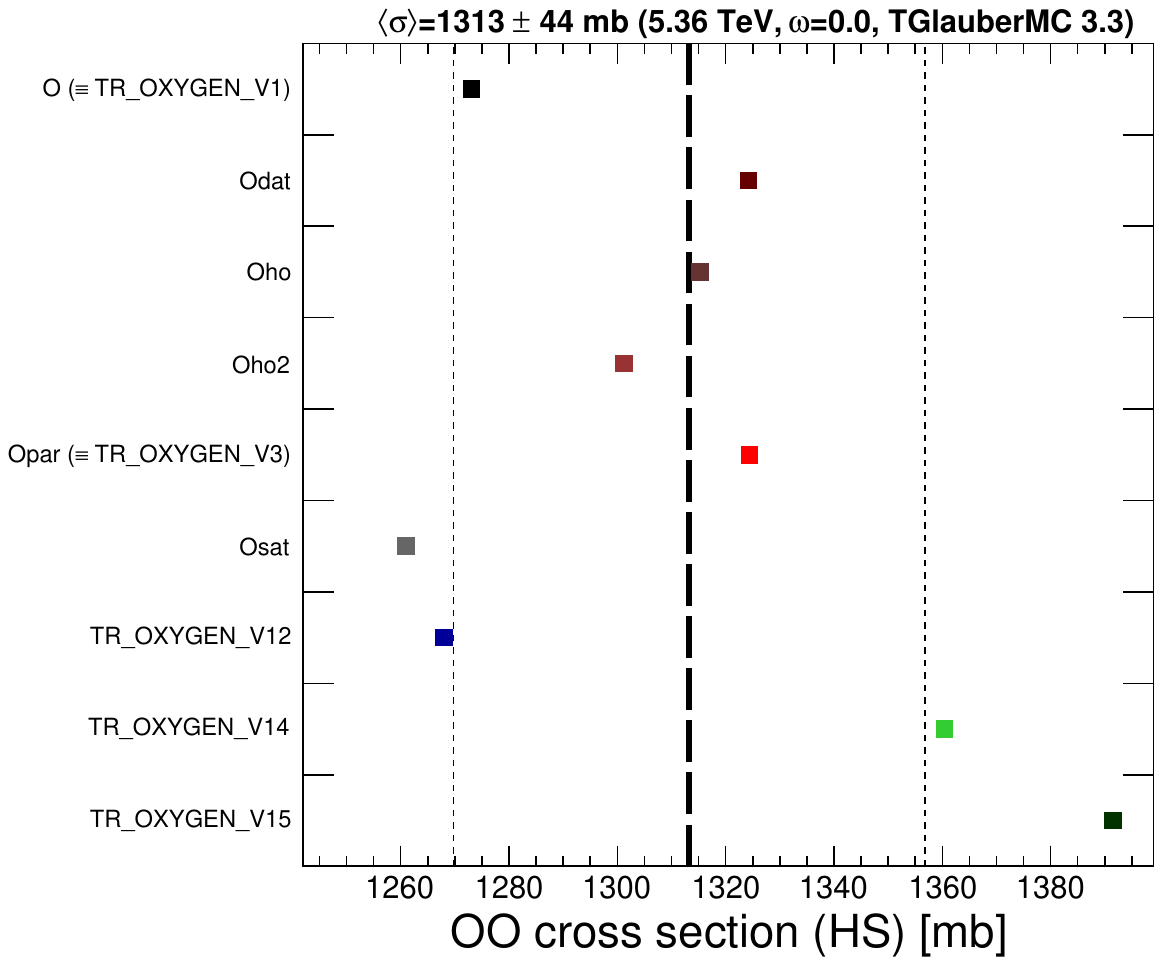}
\includegraphics[width=8cm]{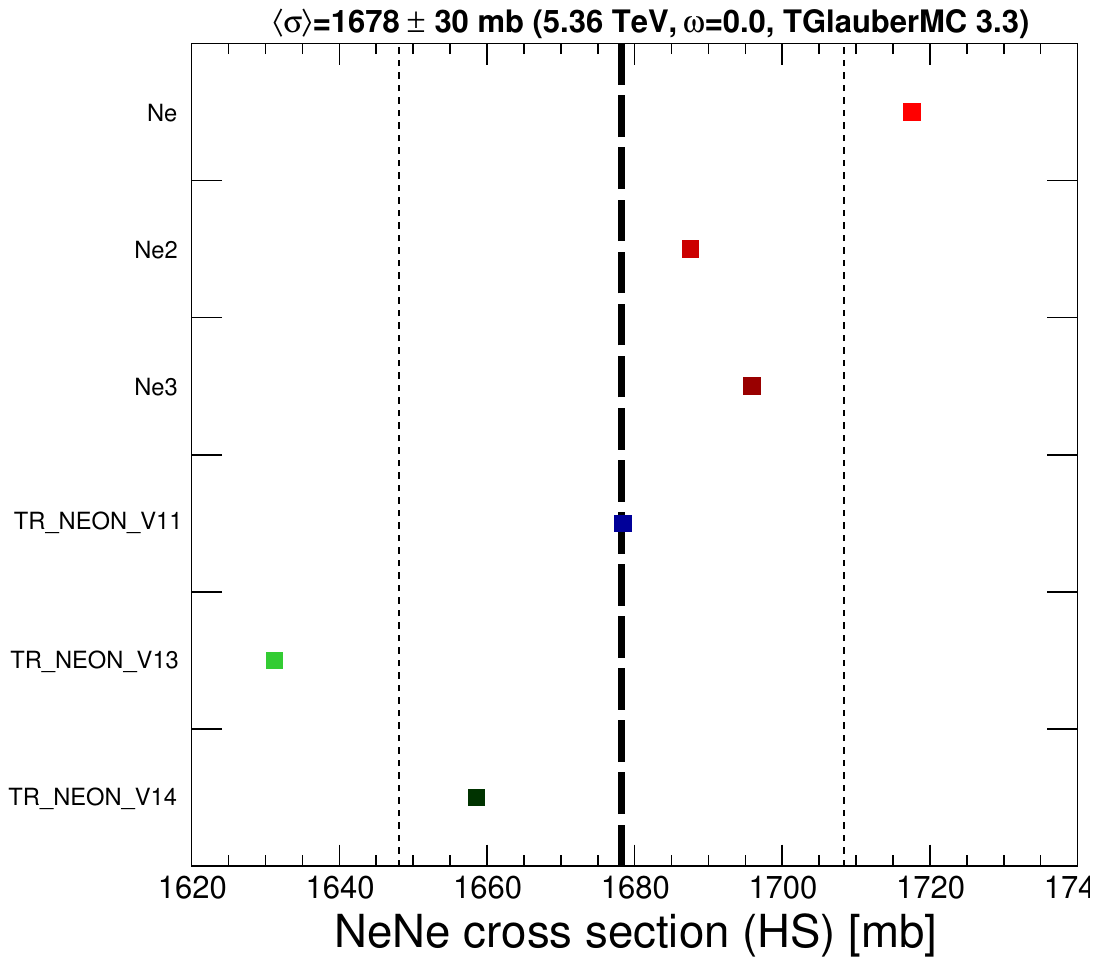}
\caption{Calculated cross sections for \OO\ collisions~(left) and \NeNe\ collisions~(right) at 5.36 TeV using TGlauberMC with $\dmin=0.4$~fm, and $\omega=0.0$~(hard-sphere approximation), using the nuclear profiles and nucleon configurations as described in \Sec{sec:nuclearprof}. The average as well as the average $\pm$ the standard deviation are shown as vertical lines.
\ifarxiv
(The same figures including available, but non-meaningful nuclear density profiles are shown in \eFig{fig:extraxsec}).
\fi
}
\label{fig:oxnecs}
\end{figure}

The resulting cross sections for \OO\ and \NeNe\ collisions at 5.36~TeV are shown in \Fig{fig:oxnecs}, for $\dmin=0.4$~fm and $\omega=0$, i.e.\ in the hard-sphere approximation, used in previous works~\cite{Loizides:2017ack,dEnterria:2020dwq}.
The cross sections averaged over all profiles discussed above are $\sigma_{\rm HS}=1313\pm44$~mb and $\sigma_{\rm HS}=1678\pm30$~mb, for \OO\ and \NeNe\ collisions, respectively.
The above uncertainties are obtained by using the nuclear profiles as described in \Sec{sec:nuclearprof}.
Instead, in case of oxygen, varying the two parameters of the ``Oho2'' model, within the reported uncertainties would results in a relative uncertainty below 1\%. 
\ifarxiv
The resulting cross section for \pO\ collisions at 9.62~TeV are shown in \eFig{fig:extraxsecpo}, leading to $\sigma_{\rm HS}=466\pm9$~mb when averaging over the various oxygen density profiles.
\fi

Recentering~(discussed above) does not affect the calculated cross section, neither does changing $\dmin$ to a value lower than 0.4~fm.
Increasing $\dmin$ beyond 0.4~fm, however starts to affect the calculated cross section~(e.g.\ for $\dmin=0.8$~fm the increase is nearly 2.5\%). 
However using $\dmin=0.8$~fm significantly distorts the charged density, which is expected given that the average minimal separation is less than 0.8~fm~(evaluated using the ab-initio nucleon configurations~\cite{Lonardoni:2017hgs} named ``O'' in TGlauberMC).
Variation of $\sigmaNN$ by $\pm1.2$~mb, leads to a change of less than 1\%.

\begin{figure}[t] 
\includegraphics[width=12cm]{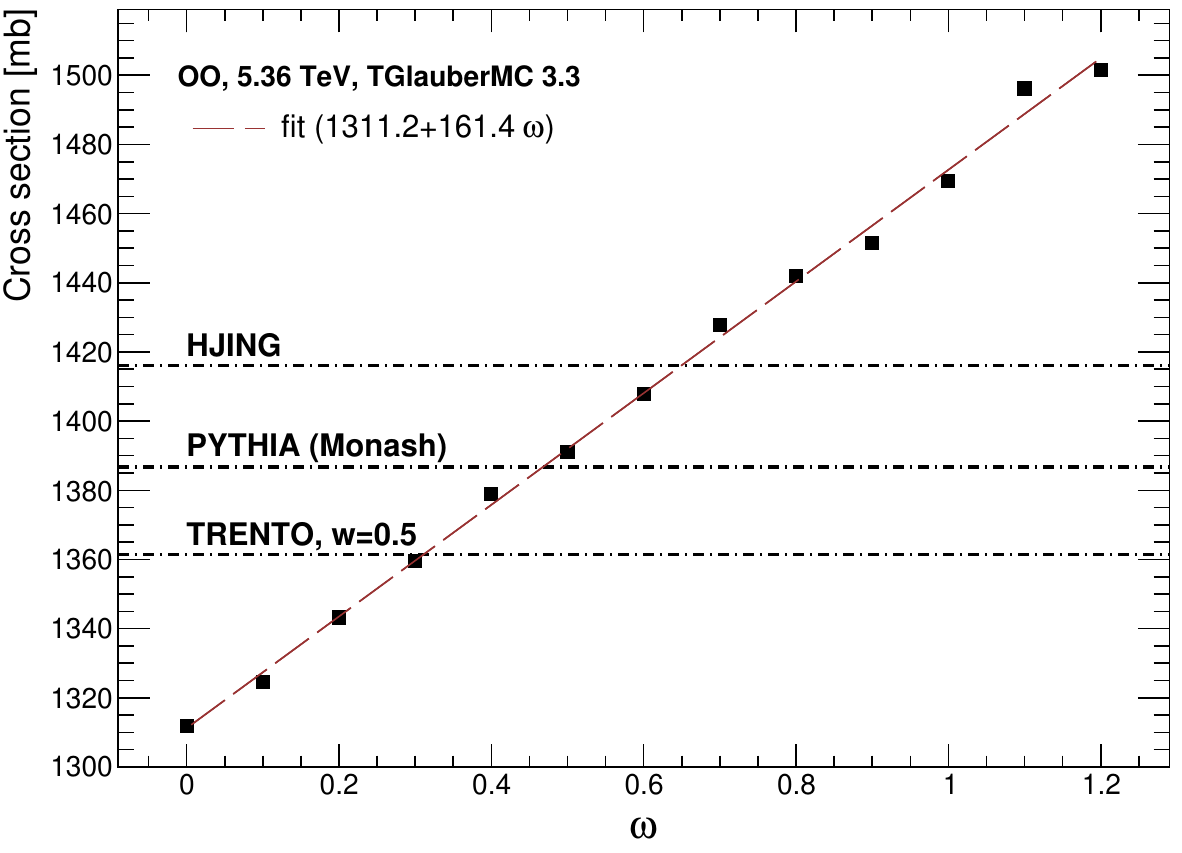}
\caption{Dependence of the calculated cross sections on $\omega$ for \OO\ collisions at 5.36 TeV using TGlauberMC with the $\Gamma$-parametrization of the nucleon--nucleon overlap profile \Eq{eq:NN_coll_profile}. 
Results using HIJING, PYTHIA~(Monash) and TRENTO with $w=0.5$ overlap profiles are given by the thin lines.}
\label{fig:ox_om_cs}
\end{figure}

The choice of $\omega=0$ is not really justified since particle production in \pp\ collisions strongly depends on the nucleon--nucleon impact parameter distribution, exemplified in the development of the multi-parton picture of the  PYTHIA~\cite{Sjostrand:2017cdm} generator to describe data obtained at hadron colliders in the TeV scale.
However, as discussed in \Sec{sec:transoverlap} there is a large parameter space beyond the hard-sphere approximation.
In particular, the calculated nucleus--nucleus cross section depends quite significantly on the nucleon--nucleon collision probability, as already discussed in \Refe{Nijs:2022rme}.
The dependence of the \OO\ cross section on $\omega$ is shown in \Fig{fig:ox_om_cs}.
The cross section is found to increase rather linearly with $\omega$ with a slope of about $161$~mb, i.e.\ the relative increase to the hard-sphere approximation is about $12.3\%\cdot\omega$.
In addition, results from HIJING, PYTHIA and TRENTO with $w=0.5$ are indicated, which roughly span the values for $\omega$ between $0.3$ and $0.6$.

\begin{figure}[t] 
\includegraphics[width=8cm]{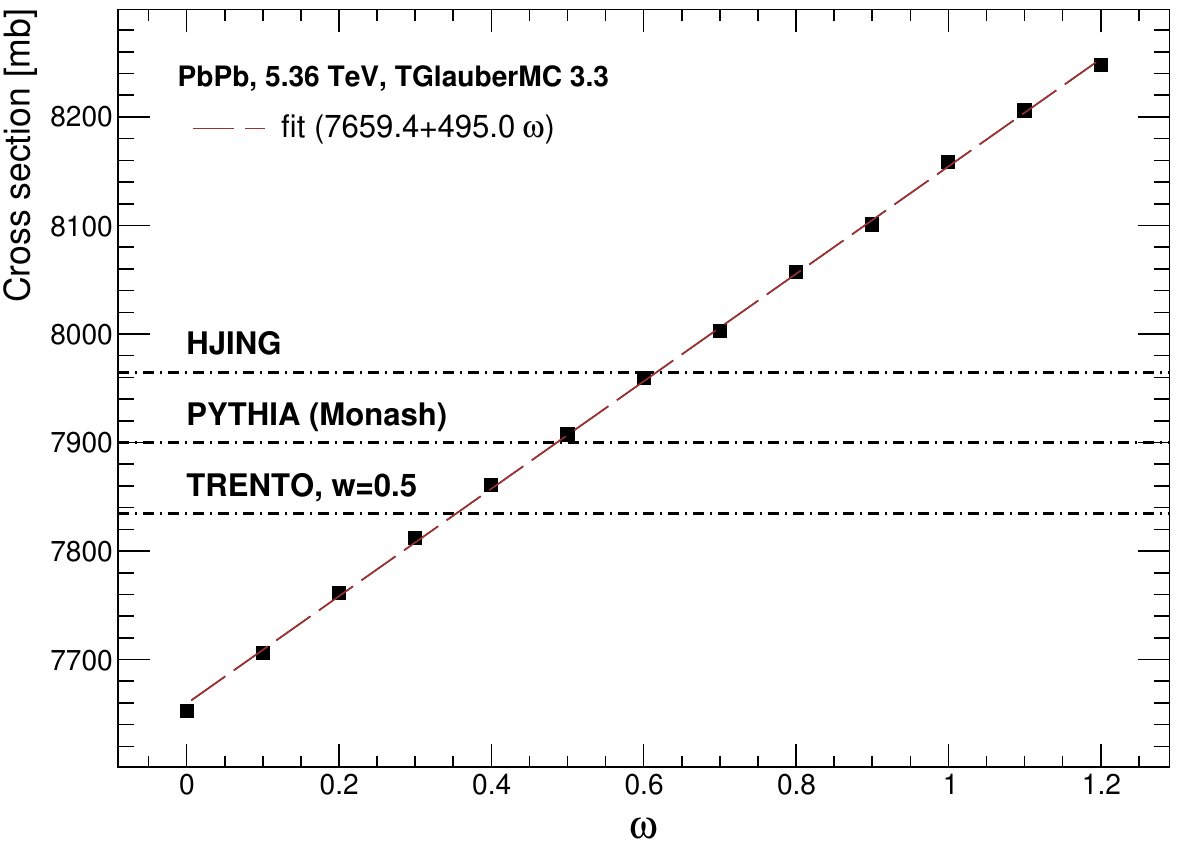} 
\includegraphics[width=8cm]{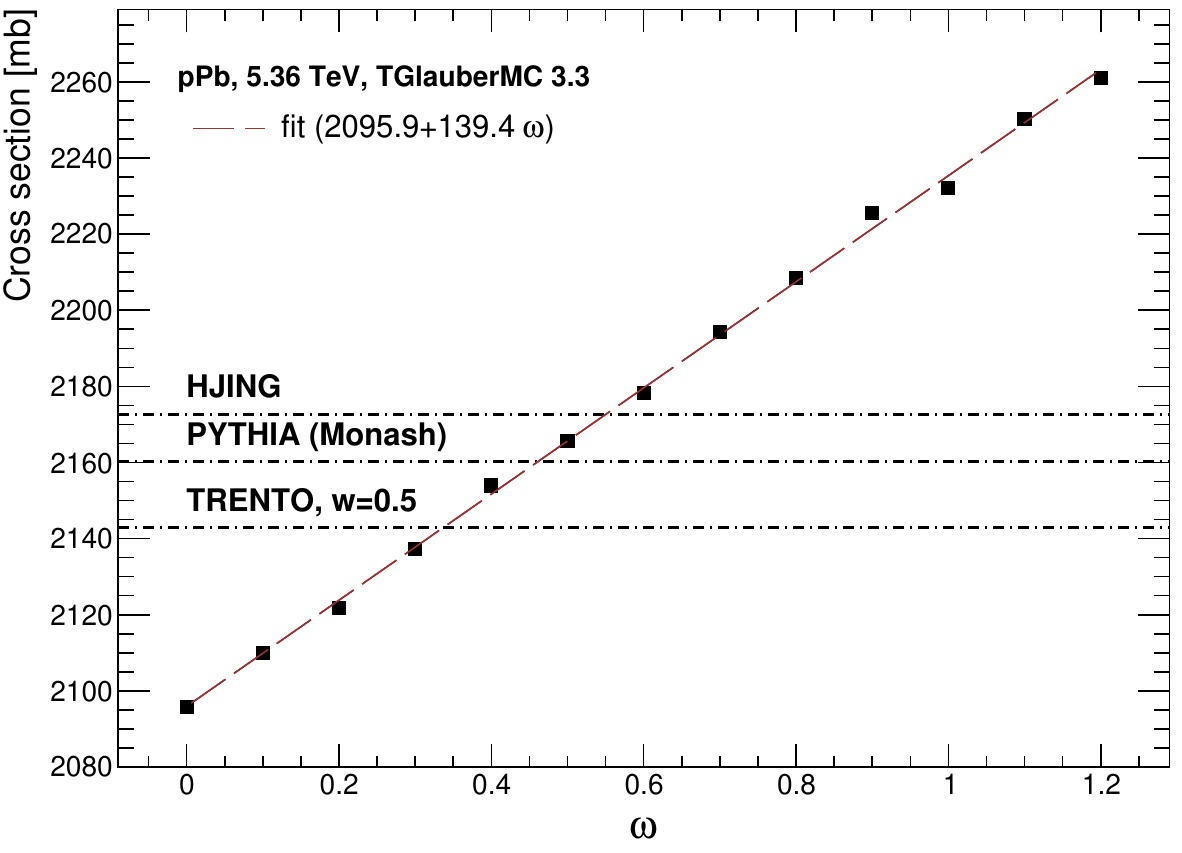} 
\includegraphics[width=8cm]{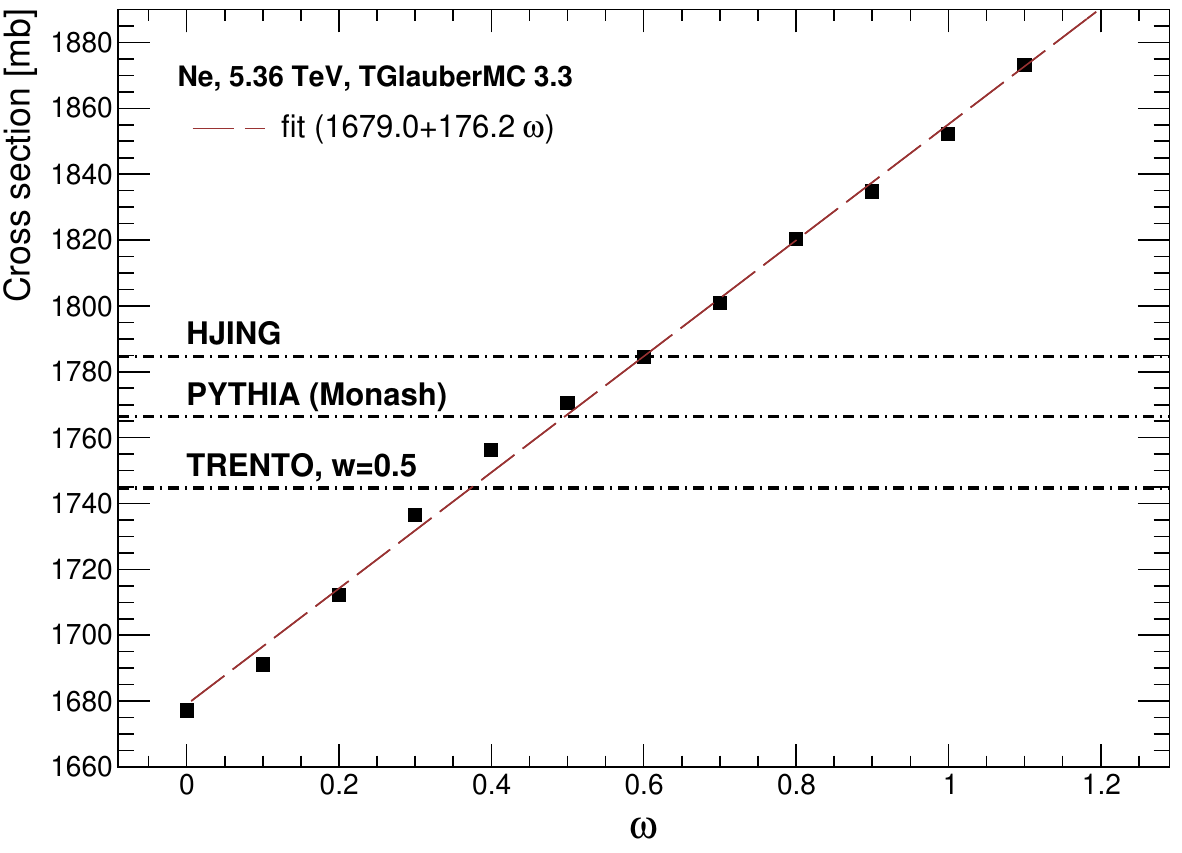}   
\includegraphics[width=8cm]{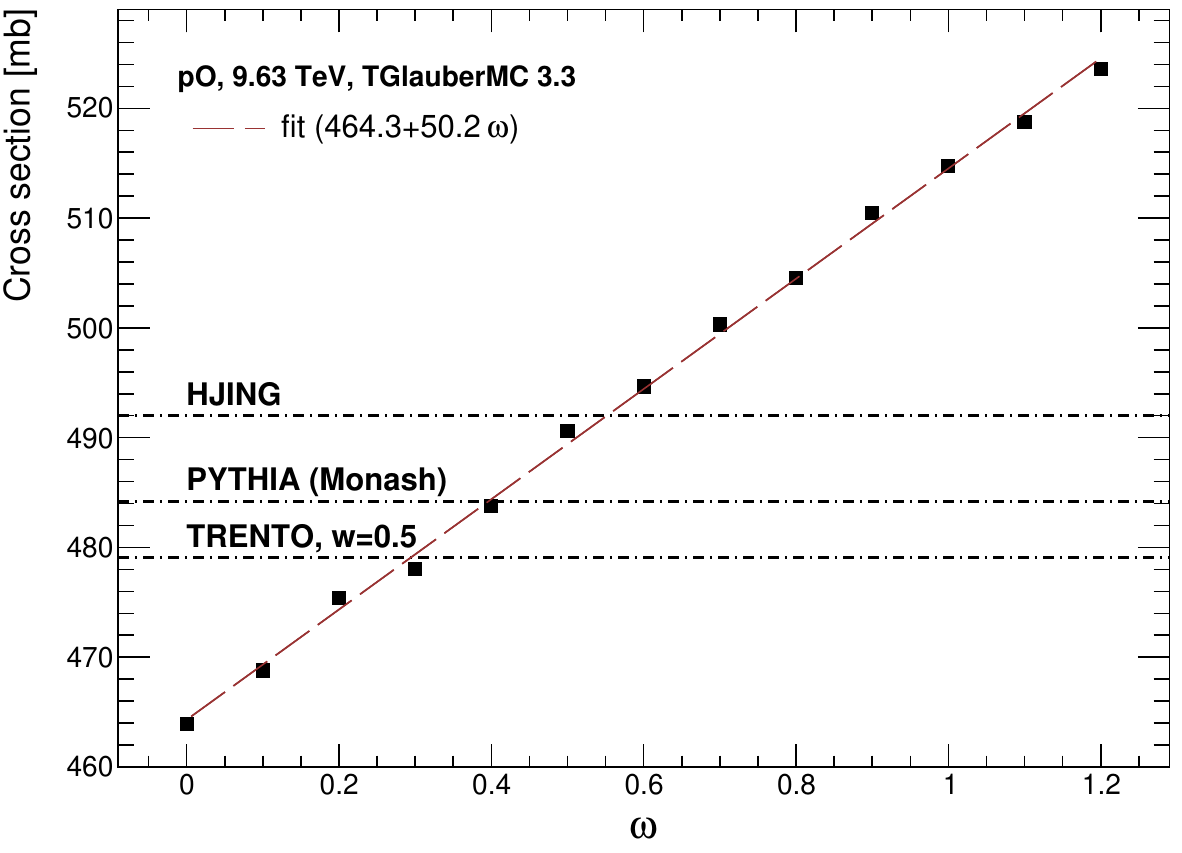}   
\caption{Dependence of the calculated cross sections on $\omega$ for \PbPb, \pPb, and \NeNe\ collisions at 5.36~TeV and \pO\ collisions at 9.62~TeV using TGlauberMC  with the $\Gamma$-parametrization of the nucleon--nucleon overlap profile \Eq{eq:NN_coll_profile}. 
Results using HIJING, PYTHIA~(Monash) and TRENTO with $w=0.5$ overlap profiles are given by the thin lines.}
\label{fig:pb_om_cs}
\end{figure}
\begin{table}[ht!b]
\centering
\caption{\label{tab:crossold} 
Measured and calculated cross sections for \PbPb\ collisions at 2.76~TeV, and \pPb\ and \PbPb\ collisions at 5.02~TeV.
The calculations using the hard-sphere approximation are from \cite{Loizides:2017ack}.
The new results using the $\Gamma$-parametrization for the nucleon--nucleon overlap function with $\omega=0.3$ are obtained in this work as motivated in the text.
} 
\vspace{0.1cm}
\begin{tabular}{l|l|l|l}
 System                 &  Measurement                                       &  Glauber (HS) \cite{Loizides:2017ack} & Glauber ($\omega=0.3$) \\\hline
 \PbPb, $\snn=2.76$~TeV & $\sigmaPbPb=7.7\pm0.6$~b~\cite{ALICE:2012aa}       & $\sigmaPbPb^{\rm HS}=7.55\pm0.15$~b   & $\sigmaPbPb^{\rm MC}=7.70\pm0.30$~b \\
 \PbPb, $\snn=5.02$~TeV & $\sigmaPbPb=7.67\pm0.25$~b~\cite{ALICE:2022xir}    & $\sigmaPbPb=7.62\pm0.15$~b            & $\sigmaPbPb^{\rm MC}=7.80\pm0.29$~b \\
 \multirow{2}{*}{\pPb, $\snn=5.02$~TeV} & $\sigmapPb=2.06\pm0.03\pm0.1$~b~\cite{CMS:2015nfb} & \multirow{2}{*}{$\sigmapPb^{\rm HS}=2.09\pm0.03$~b} & \multirow{2}{*}{$\sigmapPb^{\rm MC}=2.13\pm0.07$~b}\\ 
                                        & $\sigmapPb=2.10\pm0.07$~b~\cite{ALICE:2014gvw} & \\
\end{tabular} 
\end{table}

\begin{table}[t]
\centering
\caption{Prediction cross sections for \OO, \NeNe, \pPb\ and \PbPb\ collisions at 5.36 TeV~($\sigmaNN=68.0\pm1.2$~mb) as well as for \pO\ collisions at 9.62~TeV~($\sigmaNN=74.7\pm1.5$~mb) obtained with TGlauberMC.
Given are expected values for the hard-sphere approximation and for the case of $\omega=0.3$ for the $\Gamma$-parameterization, as well as the respective relative uncertainties. Average $\Ncoll$ values are obtained using $\Ncoll=A\,B\,\sigmaNN/\sigma_{\rm AB}^{\rm MC}$.} 
\vspace{0.1cm}
\begin{tabular}{l|l|c|l|c|c} 
 System                 &  Glauber (HS)                      & Rel.unc. & Glauber ($\omega=0.3$)               & Rel.unc. & $\langle\Ncoll\rangle$ \\\hline
\OO,   $\snn=5.36$~TeV  & $\sigmaOO^{\rm HS}=1.31\pm0.04$~b  & 3.4\%& $\sigmaOO^{\rm MC}=1.36\pm0.09$~b        & 6.8\%    & 12.8\\      
\NeNe, $\snn=5.36$~TeV  & $\sigmaNeNe^{\rm HS}=1.68\pm0.03$~b& 1.8\% & $\sigmaNeNe^{\rm MC}=1.73\pm0.08$~b    & 4.8\%     & 15.7\\
\pPb,  $\snn=5.36$~TeV  & $\sigmapPb^{\rm HS}=2.10\pm0.03$~b & 2.0\% & $\sigmapPb^{\rm MC}=2.14\pm0.07$~b     & 3.3\%     & 6.6 \\
\PbPb, $\snn=5.36$~TeV  & $\sigmaPbPb^{\rm HS}=7.66\pm0.15$  & 1.5\% & $\sigmaPbPb^{\rm MC}=7.80\pm0.30$        & 3.8\%   & 340  \\\hline
\pO,   $\snn=9.62$~TeV  & $\sigmapO^{\rm HS}=466\pm9$~mb     & 1.9\%& $\sigmapO^{\rm MC}=481\pm24$~mb           & 5.0\%   & 2.3\\
\end{tabular}
\label{tab:xsres}
\end{table}

Using an overlap function different from the hard-sphere approximation affects also \PbPb\ and \pPb\ collisions, as shown in \Fig{fig:pb_om_cs} at 5.36~TeV, with a slope of about 495 and 139~mb, respectively.
The relative increase is the same for the two systems, about 6.5\%.
For completeness, also \NeNe\ collisions at 5.36~TeV, and \pO\ collisions at 9.62~TeV are shown in \Fig{fig:pb_om_cs}, with slopes of 1679 and 464~mb, respectively.

To help deciding on a value for $\omega$, existing cross section data at the LHC are provided in \Tab{tab:crossold} together with earlier calculations using the hard-sphere approximation~\cite{Loizides:2017ack}.
As can be seen, there is not a lot of room to increase the cross section beyond the hard-sphere approximation.
Quite the contrary, the calculated cross sections are almost perfectly agreeing with the measurements.
The recent Bayesian analysis~\cite{Nijs:2022rme} mentioned earlier, which includes a wealth of data points different from cross sections, obtained $0.4<w<0.5$ for the TRENTO model, which is similar to $\omega=0.3$ case~(see \Fig{fig:ox_om_cs} and \Fig{fig:pb_om_cs}).
Indeed, the $\chi^2/n_{\rm dof}$ between the available data points and the $\omega$-dependent cross section obtained from the linear fits is about $0.6$ for $\omega = 0.3$, and the corresponding increase is comparable in size to the systematic uncertainty quoted for the hard-sphere–based calculation.
It hence is reasonable to increase the predicted cross sections by the associated relative amount (between 1.5\% and 2\%), and assign the same value as additional uncertainty. 
In this way, also the HIJING overlap profile, which qualitatively describes the centrality dependence of particle production in many collision systems~\cite{ALICE:2014xsp,Loizides:2017sqq,ALICE:2018ekf,dEnterria:2020dwq}, is adequately covered.
Using this prescription, the predicted cross sections for $\omega=0.3$ are given in \Tab{tab:crossold}, together with the values using the hard-sphere approximation.

\section{Initial-condition related distributions }
\label{sec:init}

\begin{figure}[t] 
\includegraphics[width=8cm,height=5.8cm]{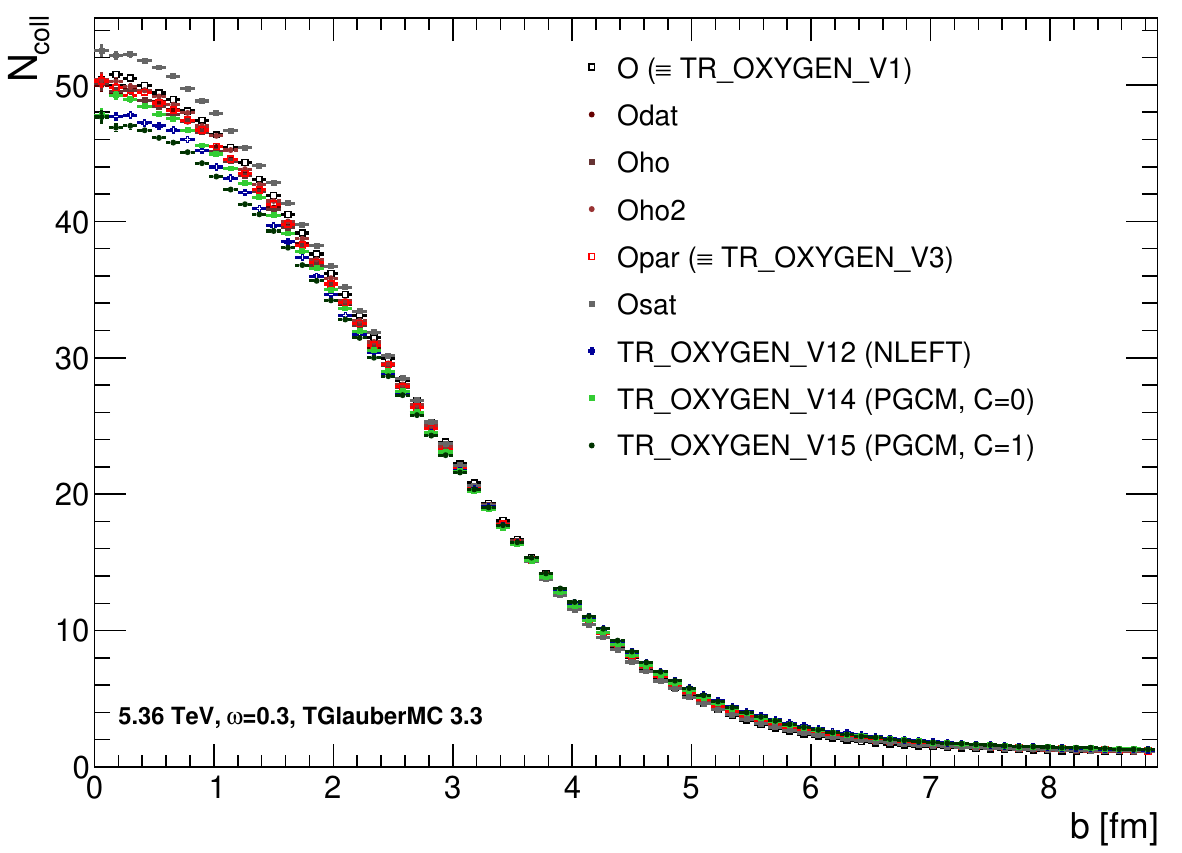}
\includegraphics[width=8cm,height=5.8cm]{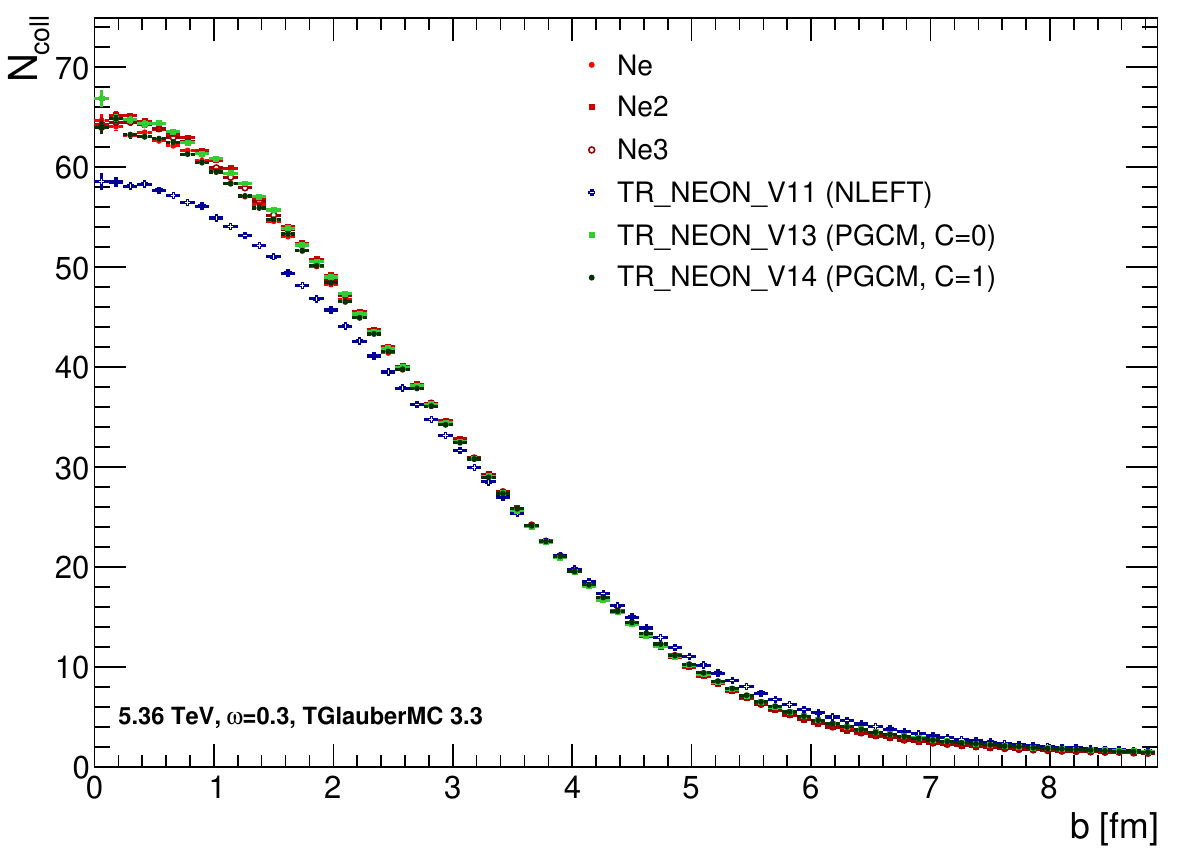}

\vspace{-0.56cm}
\includegraphics[width=8cm,height=2.9cm]{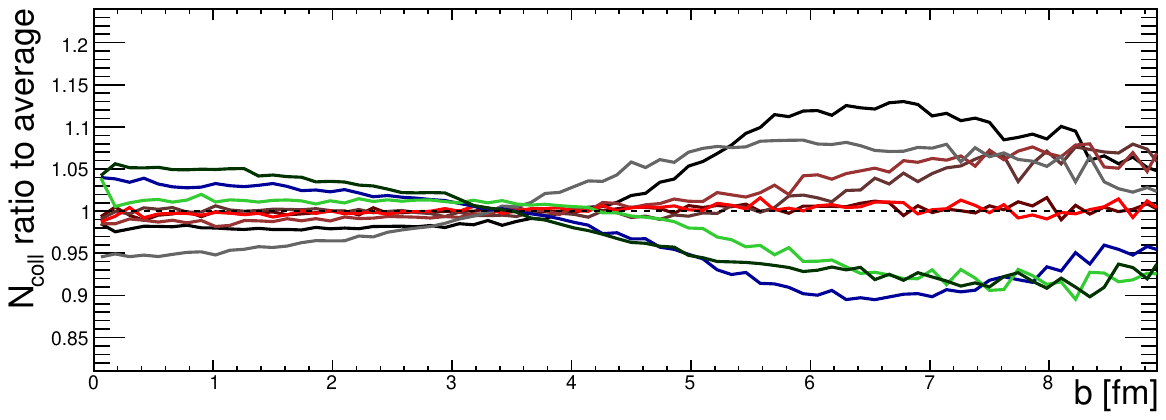}
\includegraphics[width=8cm,height=2.9cm]{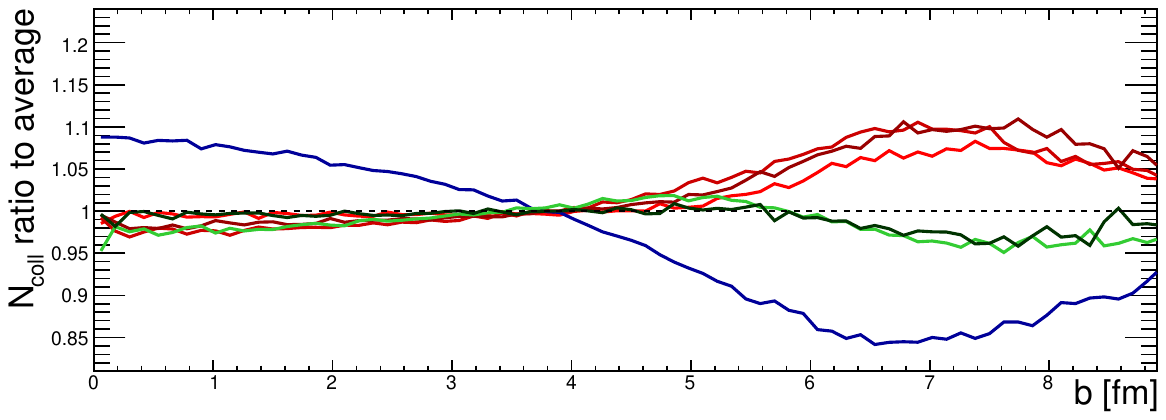}

\vspace{0.2cm}
\includegraphics[width=8cm,height=5.8cm]{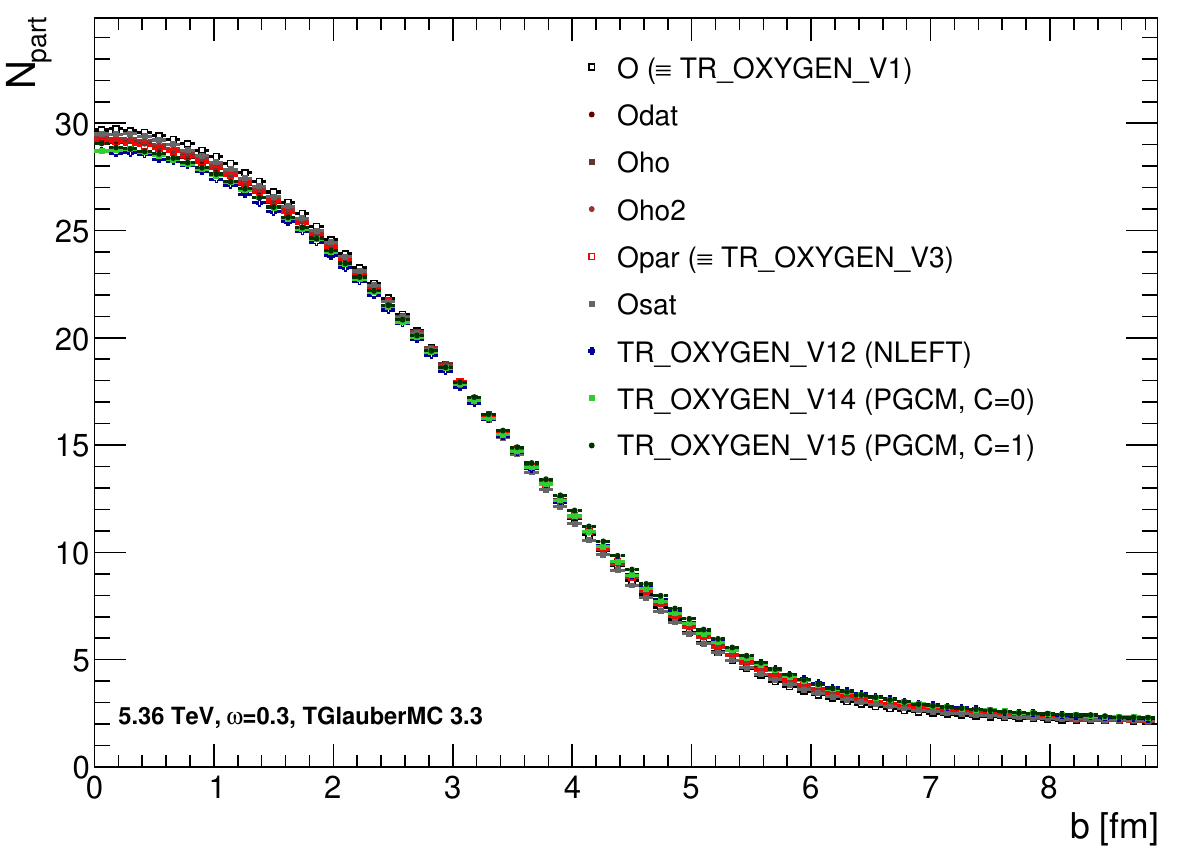}
\includegraphics[width=8cm,height=5.8cm]{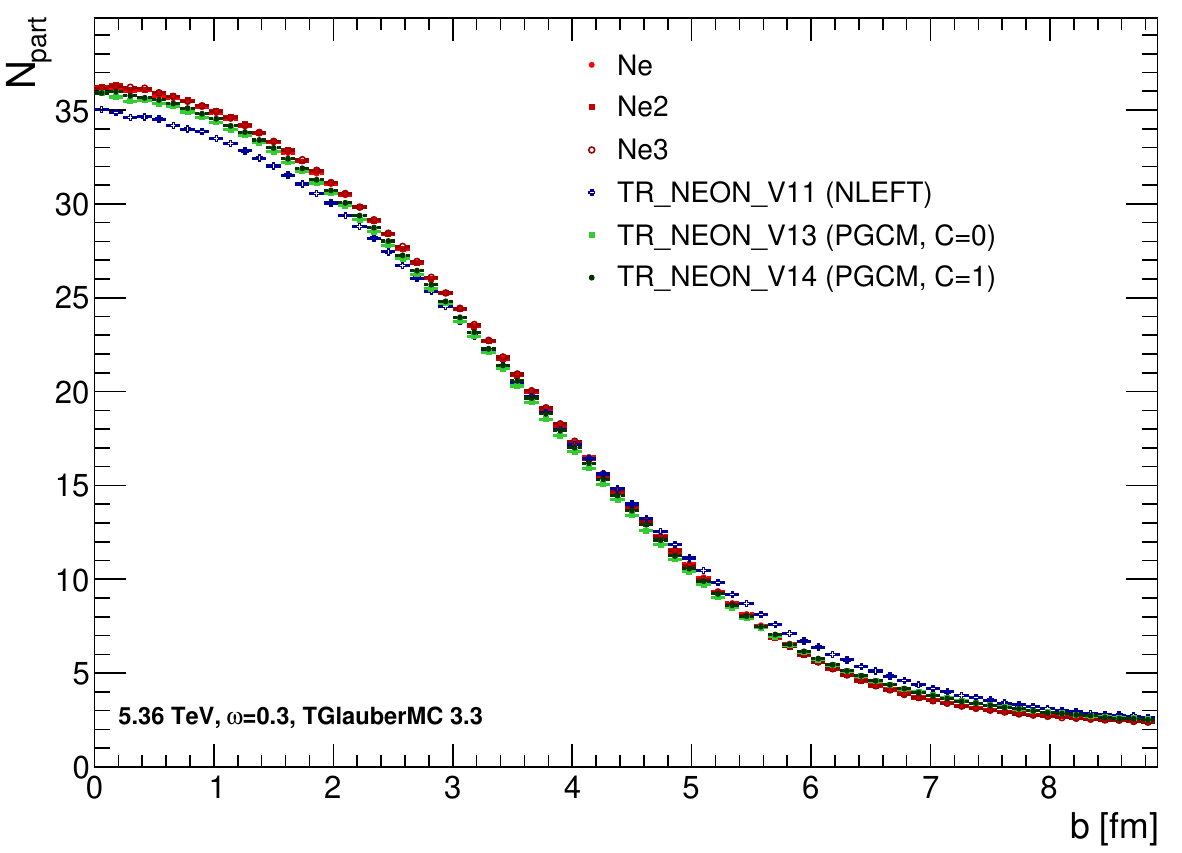}

\vspace{-0.56cm}
\includegraphics[width=8cm,height=2.9cm]{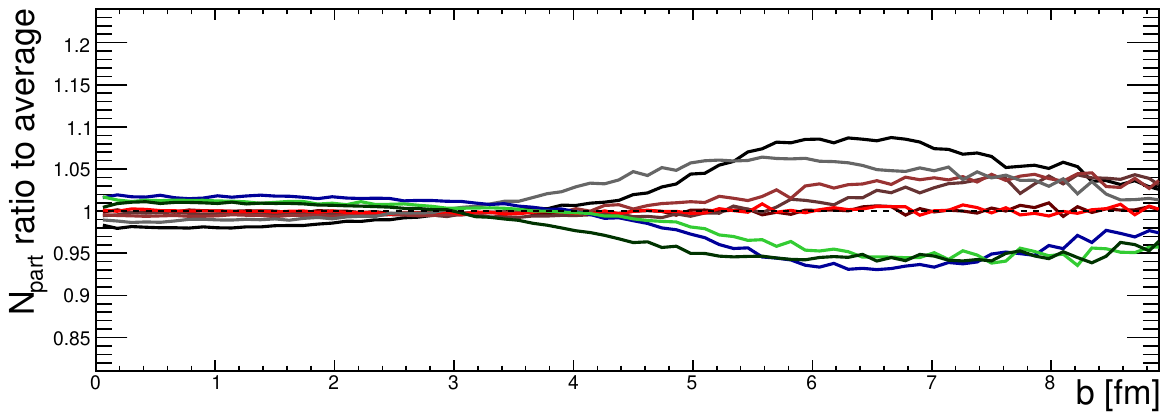}
\includegraphics[width=8cm,height=2.9cm]{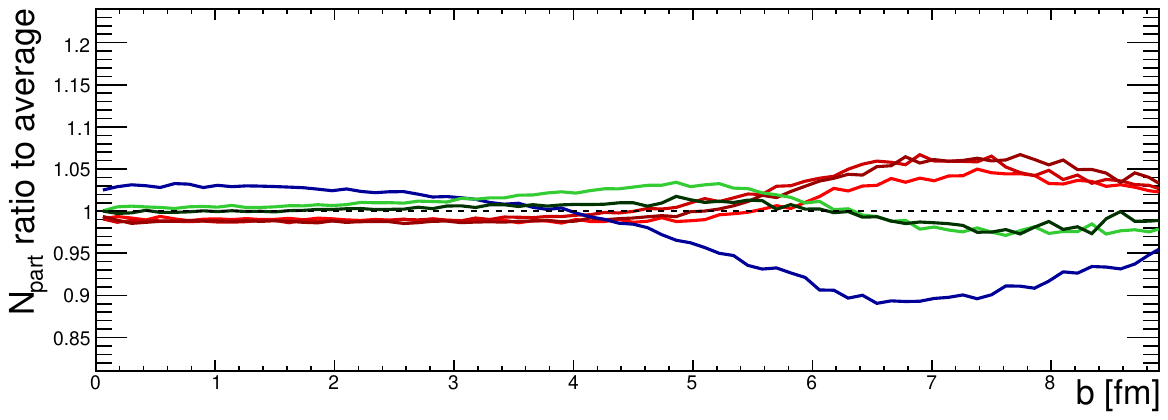}
\caption{Number of collisions~(top row) and number of participants~(bottom row) versus impact parameter together for \OO~(left column) and \NeNe~(right column) collisions at 5.36~TeV using TGlauberMC with $\omega=0.3$ for the nuclear density profiles discussed in \Sec{sec:nuclearprof}. The smaller bottom panels show the respective ratio to the average over all density profiles.}
\label{fig:quantities}
\end{figure}

\begin{figure}[t] 
\includegraphics[width=8cm,height=5.8cm]{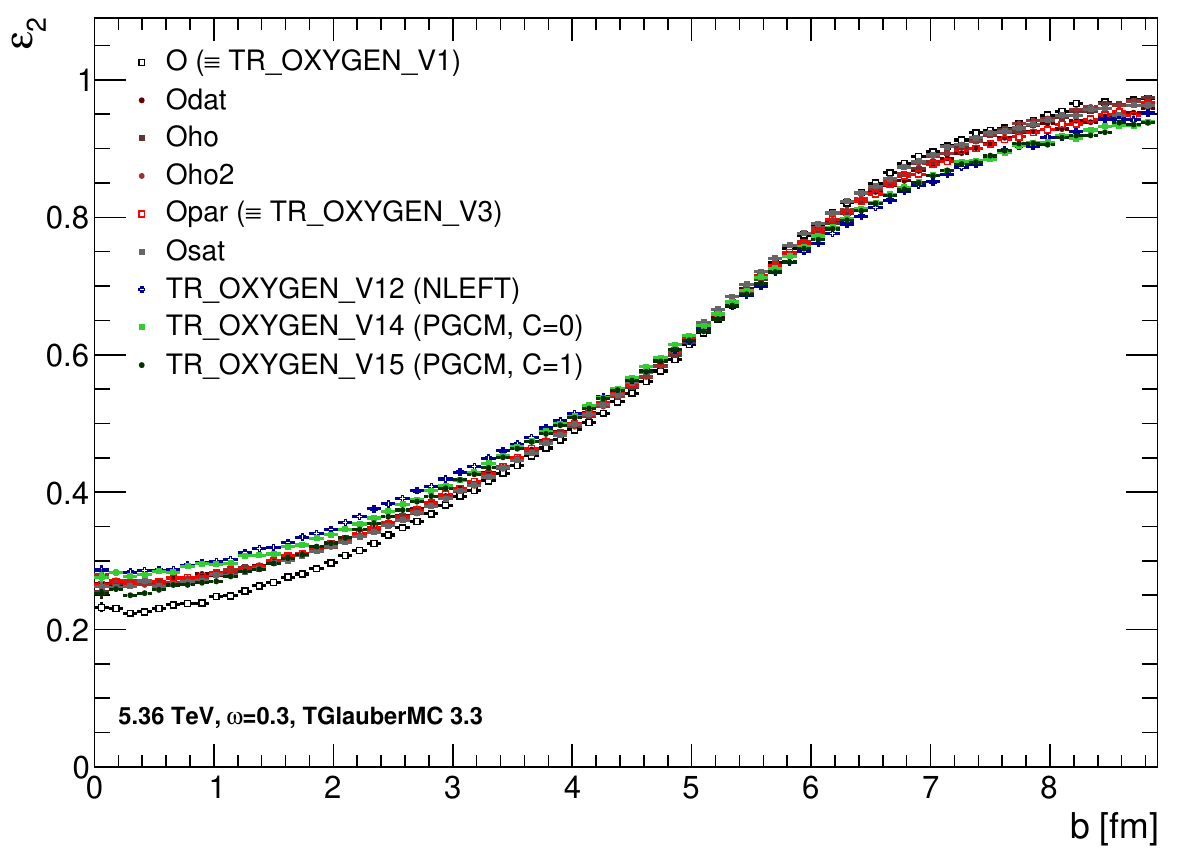}
\includegraphics[width=8cm,height=5.8cm]{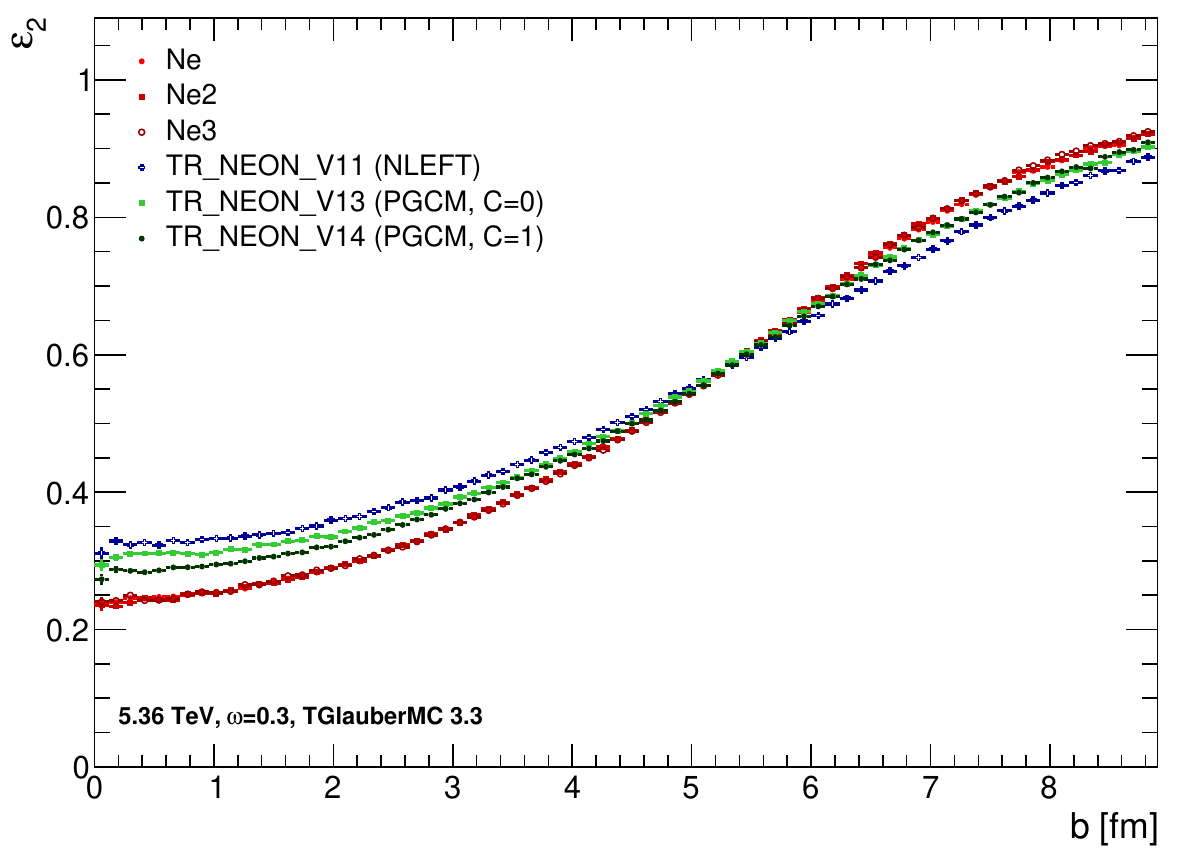}

\vspace{-0.56cm}
\includegraphics[width=8cm,height=2.9cm]{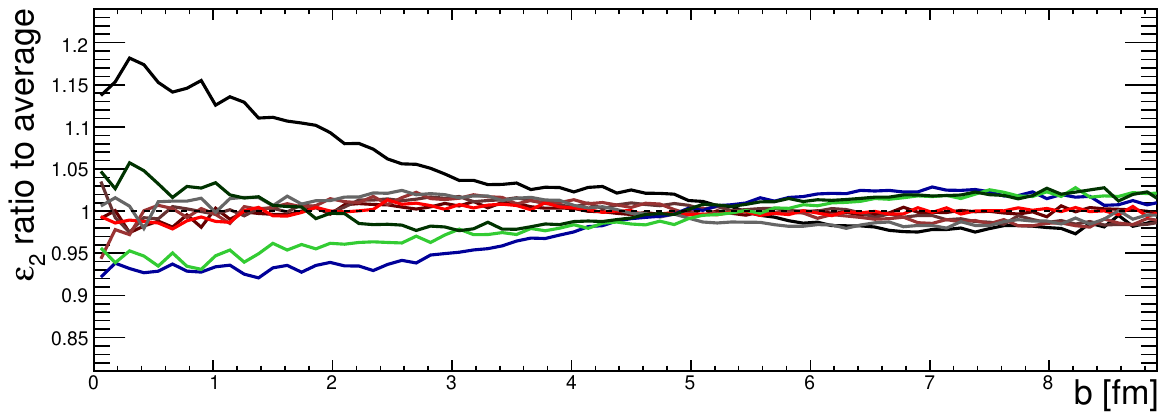}
\includegraphics[width=8cm,height=2.9cm]{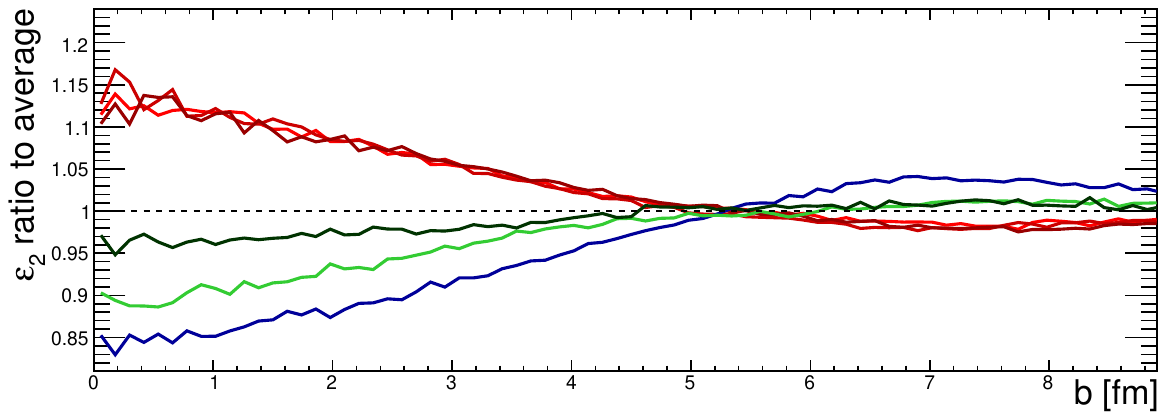}
\caption{Participant eccentricity versus impact parameter~(bottom row) for \OO~(left column) and \NeNe~(right column) collisions at 5.36~TeV using TGlauberMC with $\omega=0.3$ for the nuclear density profiles discussed in \Sec{sec:nuclearprof}. The smaller bottom panels show the respective ratio to the average over all density profiles.}
\label{fig:quantities2}
\end{figure}
\begin{figure}[ht!] 
\includegraphics[width=10.5cm]{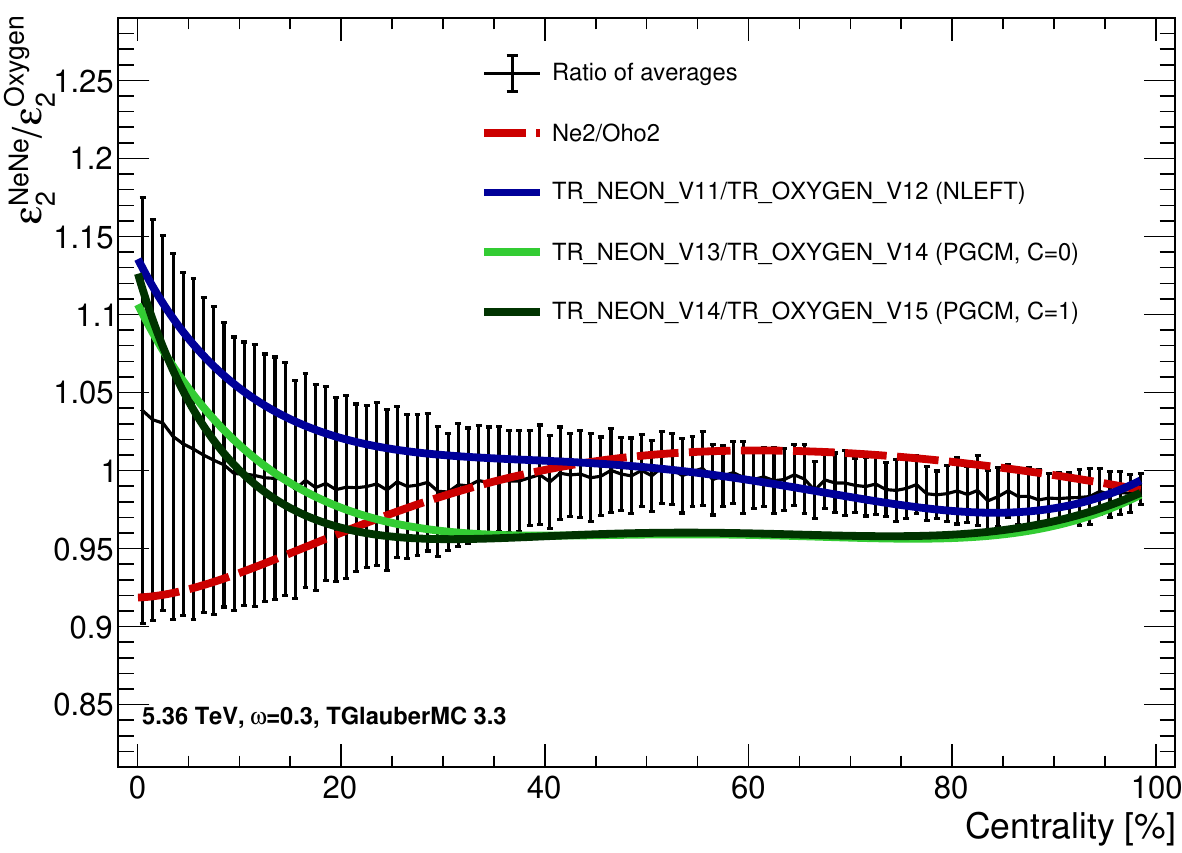}
\caption{Ratio of participant eccentricity versus centrality for \NeNe\ collisions to that for \OO\ collisions at 5.36~TeV using TGlauberMC with $\omega=0.3$ for the ratio of the averages~(standard deviation is shown as uncertainty per point), example ratio using ``Ne2'' and ``Oho2'' parameterizations as well as the ratios obtained using the nuclear densities listed in \Tab{tab:troxygen}.
}
\label{fig:v2moneyplot}
\end{figure}

\begin{figure}[ht!] 
\includegraphics[width=8cm]{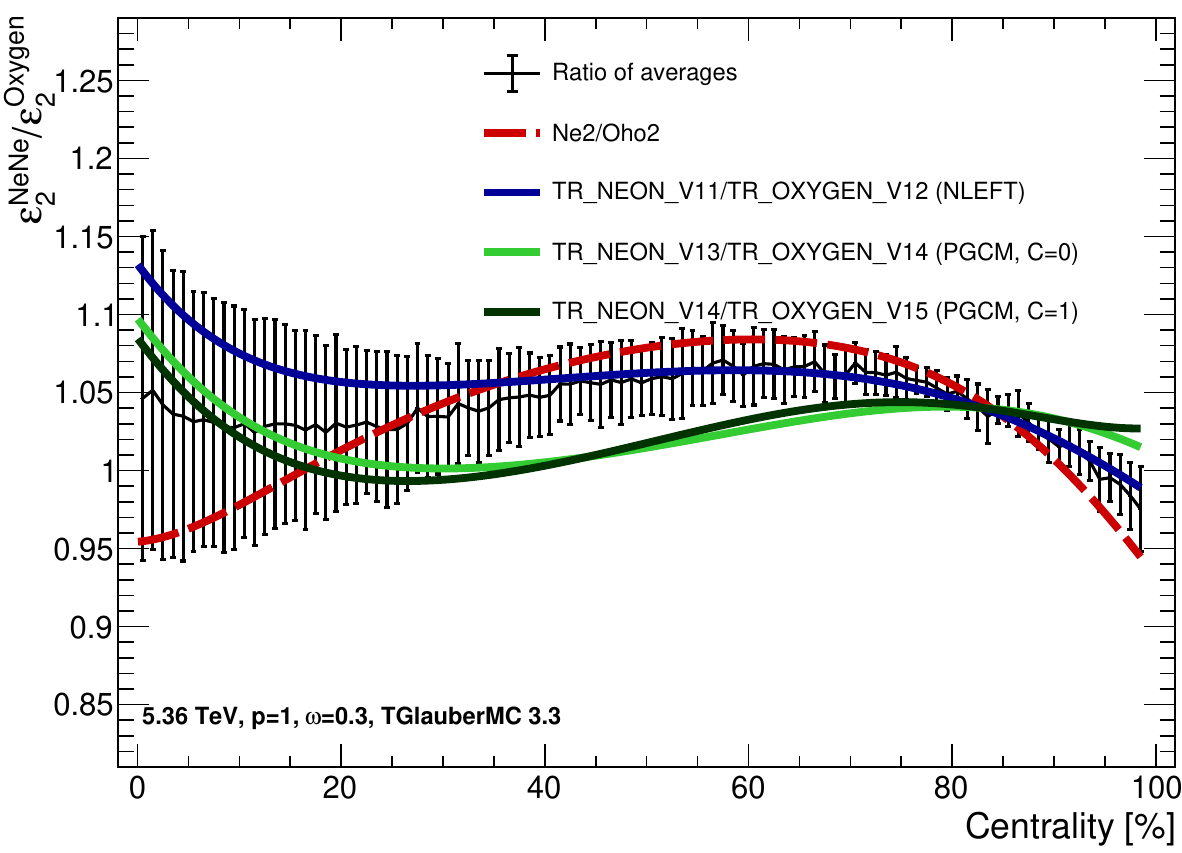}
\includegraphics[width=8cm]{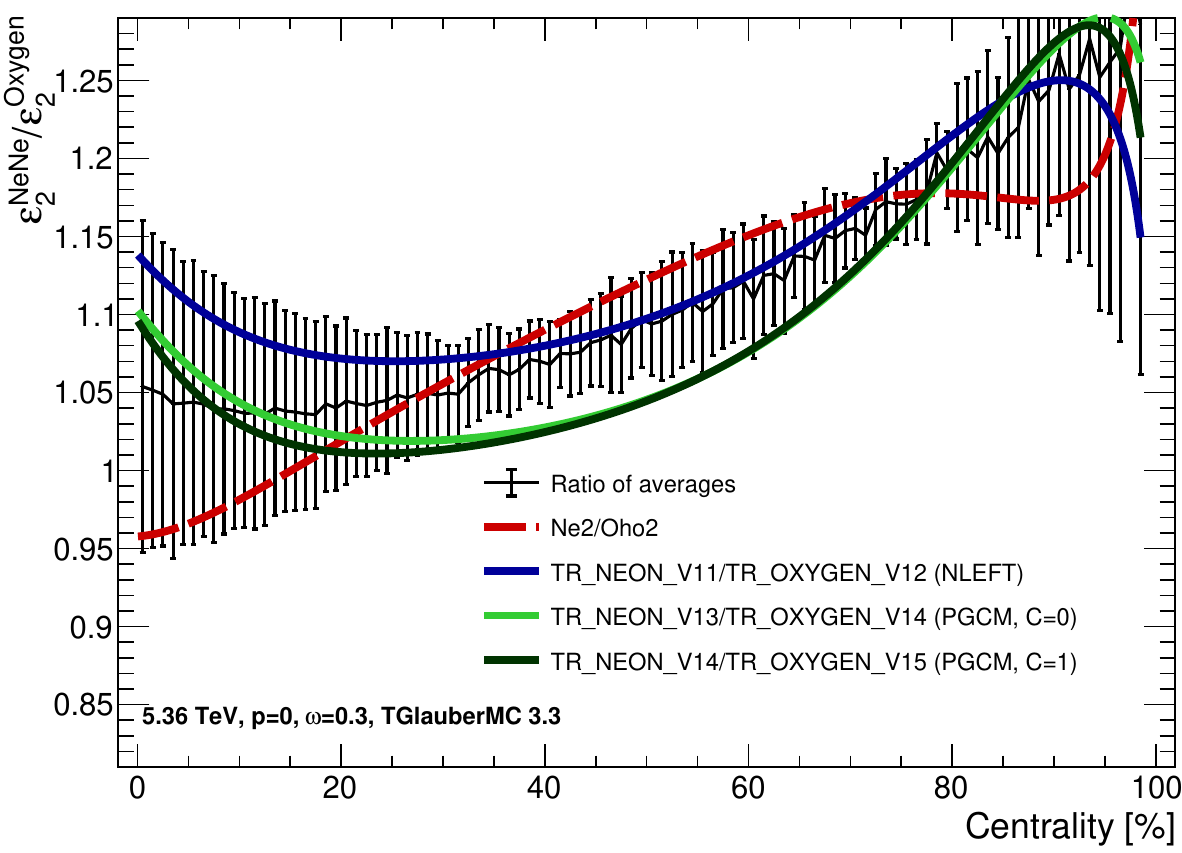}
\caption{
\ifarxiv
Ratio of participant eccentricity versus centrality for \NeNe\ collisions to that for \OO\ collisions at 5.36~TeV using TGlauberMC with $\omega=0.3$ for the ratio of the averages~(standard deviation is shown as uncertainty per point), example ratio using ``Ne2'' and ``Oho2'' parameterizations as well as the ratios obtained using the nuclear densities listed in \Tab{tab:troxygen}.
Gaussian smearing with $p=1$~(left) and $p=0$~(right panel) for the nuclear overlap is used as explained in the text.
\fi
}
\label{fig:v2moneyplotsmeared}
\end{figure}

In the following, the number of participants~($\Npart$) typically needed to compare measured quantities across systems and number of collisions~($\Ncoll$) typically used to quantify the nuclear modification factor are discussed.
Average $\Ncoll$ can be obtained using $\Ncoll=A\,B\,\sigmaNN/\sigma_{\rm AB}^{\rm MC}$ leading to $12.8\pm0.8$, $15.8\pm1.1$, $2.49\pm0.12$ for \OO\ and \NeNe\ at 5.36~TeV and \pO\ at 9.62~TeV~(see \Tab{tab:xsres}).
In case of \pO\ (or \pPb) collisions, it follows that average $\Npart=\langle\Ncoll\rangle+1$.

\Figure{fig:quantities} shows the $\Ncoll$ and $\Npart$ distributions versus the impact parameter $b$ for \OO\ and \NeNe\ collisions at 5.36~TeV computed with TGlauberMC~($\dmin=0.4$fm, $\omega=0.3$) using the nuclear density profiles discussed in \Sec{sec:nuclearprof}.
In more central collisions~(smaller impact parameter) the \Ncoll\ values differ with respect to the average by 5\% for oxygen and up 10\% for neon.
In peripheral collisions~(larger impact parameter) the differences increase to 10\% and 15\%, respectively, and a grouping into different types of underlying
nuclear density profiles~(Wood-Saxon, NLEFT, PGCM) becomes apparent.
For \Npart\ the differences are smaller, on the level of few percent in central collisions, reaching 5\% for oxygen and 10\% for neon in peripheral collisions.
In the actual measurements, the observables will be measured in intervals of multiplicity and/or energy, which for large nuclei are well correlated with centrality.
The intervals then will be mapped to $\Npart$~(and $\Ncoll$) with different procedures summarized in \Refe{Miller:2007ri,dEnterria:2020dwq}, which rely on \MCG\ calculations.
The process requires the data and calculations to be matched at the so-called anchor point. 
In \PbPb\ collisions the anchor point can be determined rather precisely to the 1\% level, and up to 90\% peripheral events~\cite{ALICE:2013hur}.
For \OO~(and \NeNe) collisions, this may not be achievable, due to the much larger fluctuations and the unknowns of the underlying nuclear density profile.

\Figure{fig:quantities2} shows the participant eccentricity versus the impact parameter $b$ for \OO\ and \NeNe\ collisions at 5.36~TeV.
Both exhibit an increasing trend from central to peripheral collisions reflecting the increasing elliptical shape of the overlap region.
Due the significant differences in the nuclear profiles, there are differences of up to 15\% relative to the average over all profiles, in particular in central collisions.

\Figure{fig:v2moneyplot} shows the ratio of participant eccentricity for \NeNe\ collisions to that for \OO\ collisions at 5.36~TeV versus centrality (estimated from the respective impact parameter distributions) for the ratio of the averages (including the standard deviation as uncertainty) as well as direct ratios from the nuclear densities listed in \Tab{tab:troxygen}.
As can be seen, the average ratio rises with increasing impact parameter, reaching about 1.05 in central collision.
On the other hand, the NLEFT and PGCM profiles reach values significantly larger values up to 1.10--1.15.
Assuming an hydrodynamical expansion of a nearly inviscid fluid, $\varepsilon_{2} \propto v_{2}$ holds to first order~\cite{Noronha-Hostler:2015dbi}, the ratio of $v_2$ between oxygen and neon is predicted to be of similar size.
Indeed, a detailed calculation~\cite{Giacalone:2024luz} expects the ratio to be $1.15$ to within $\pm3\%$ for most central collisions, since most theoretical uncertainties cancel out in the $v_2$ ratio.

\begin{figure}[t] 
\includegraphics[width=8cm,height=5.8cm]{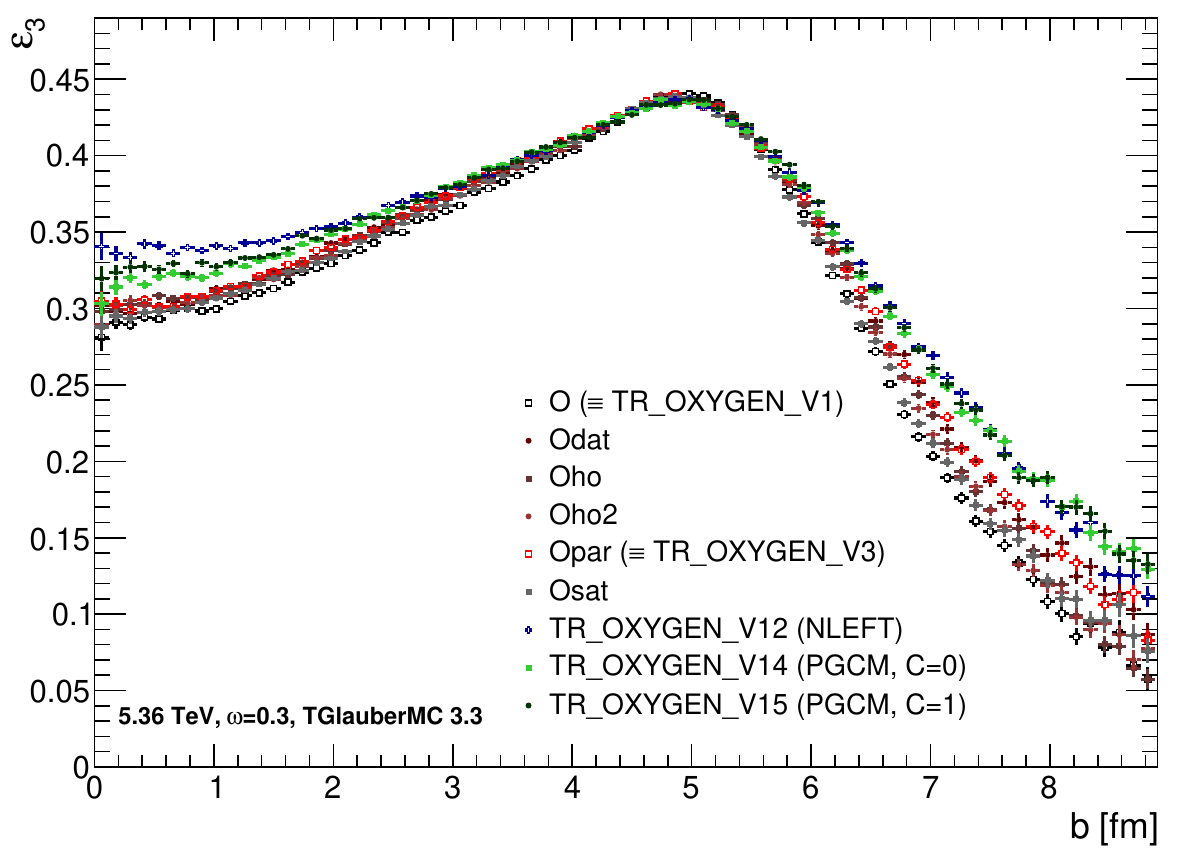}
\includegraphics[width=8cm,height=5.8cm]{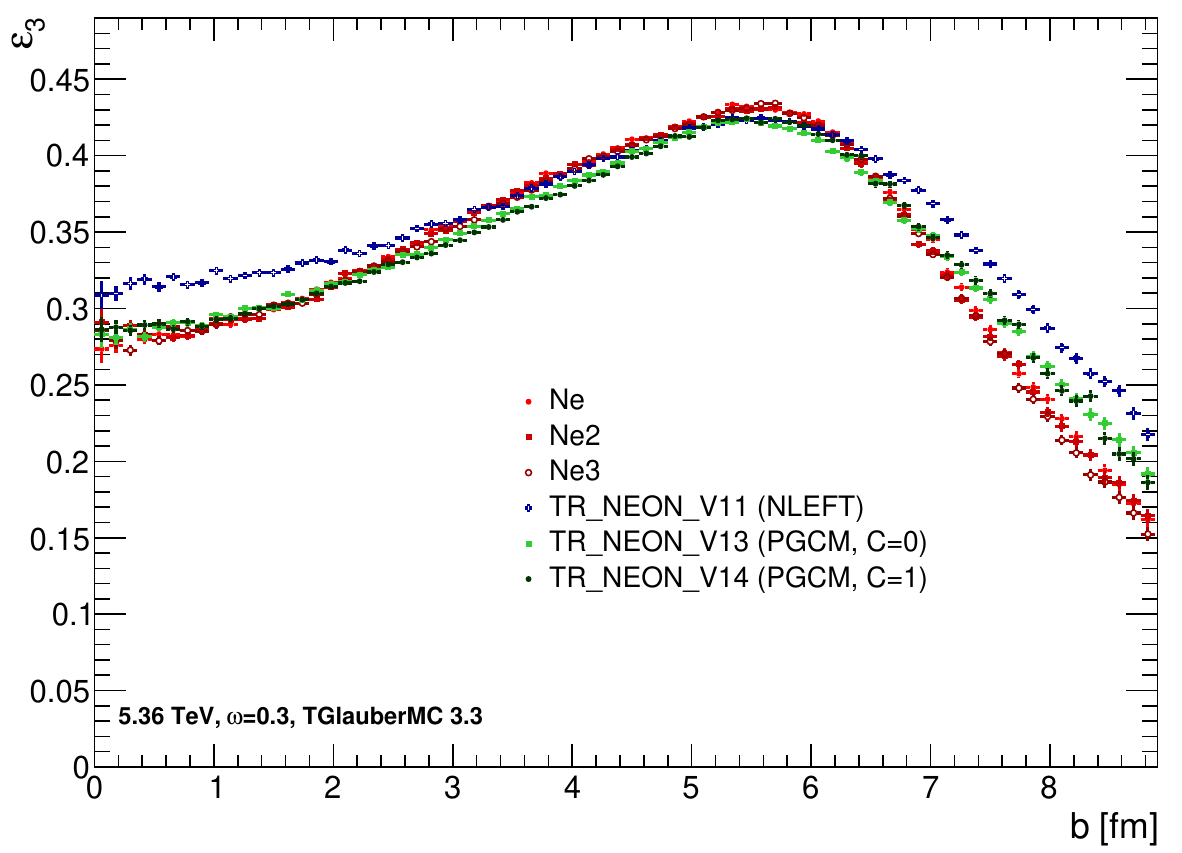}

\vspace{-0.56cm}
\includegraphics[width=8cm,height=2.9cm]{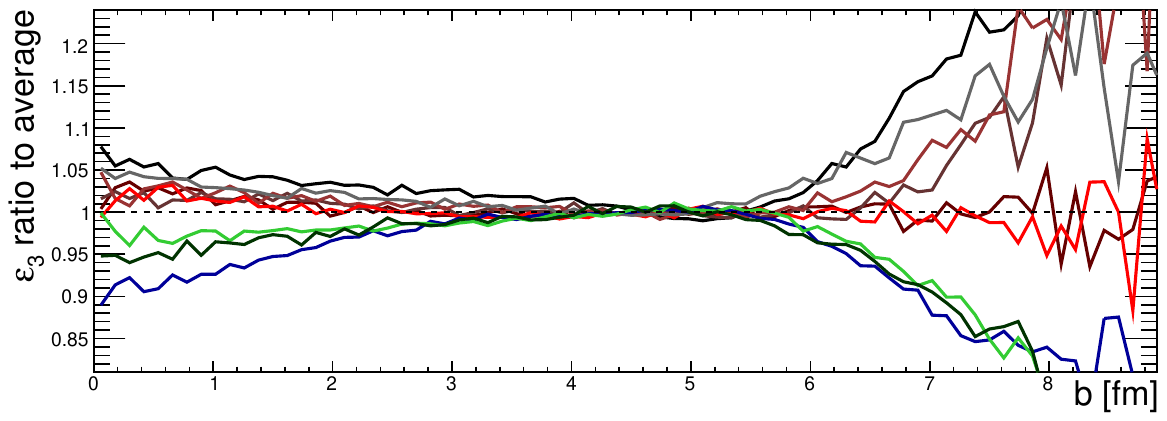}
\includegraphics[width=8cm,height=2.9cm]{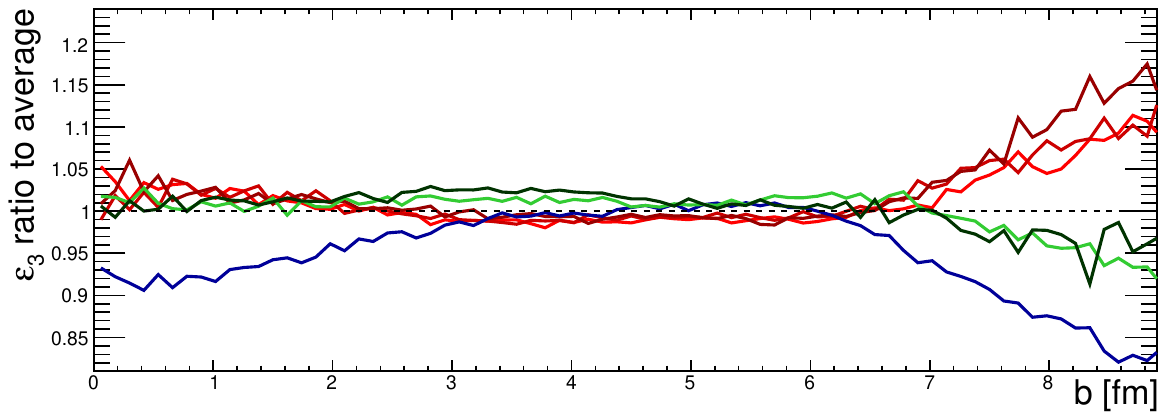}

\vspace{0.2cm}
\includegraphics[width=8cm,height=5.8cm]{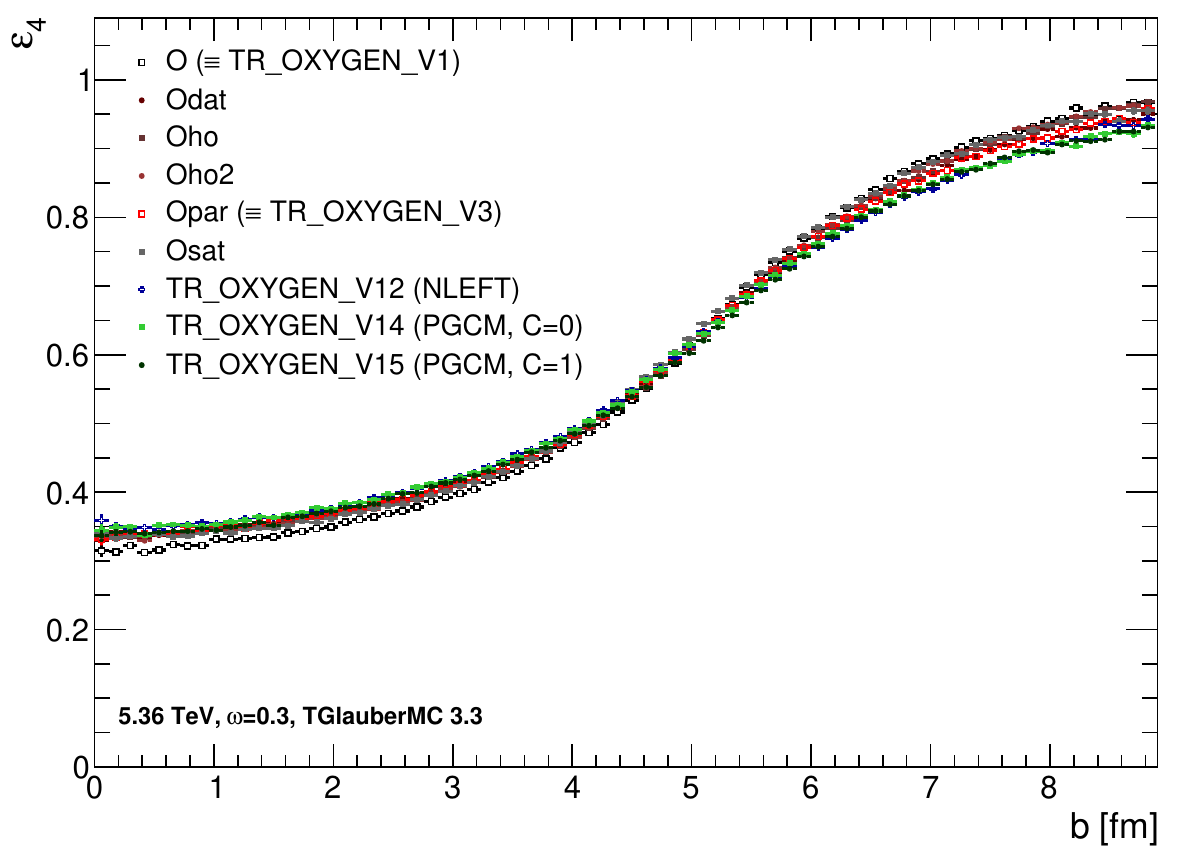}
\includegraphics[width=8cm,height=5.8cm]{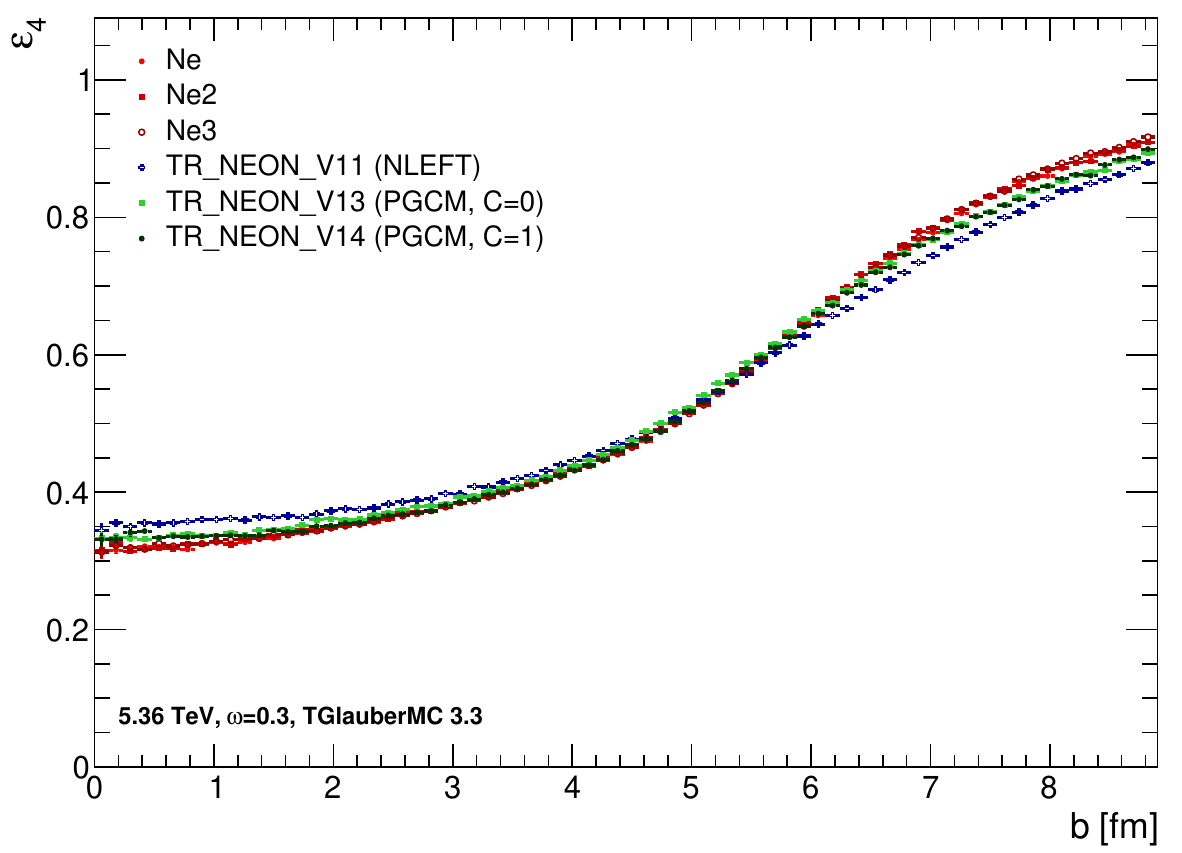}

\vspace{-0.56cm}
\includegraphics[width=8cm,height=2.9cm]{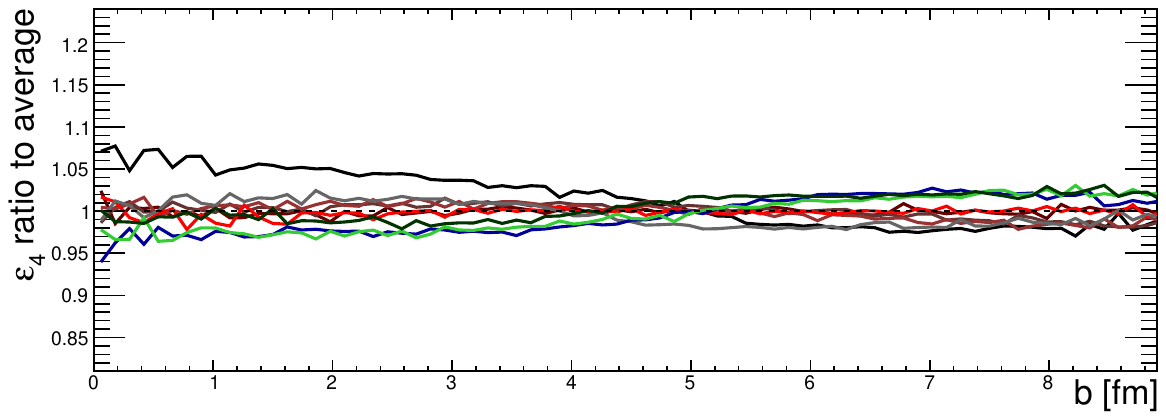}
\includegraphics[width=8cm,height=2.9cm]{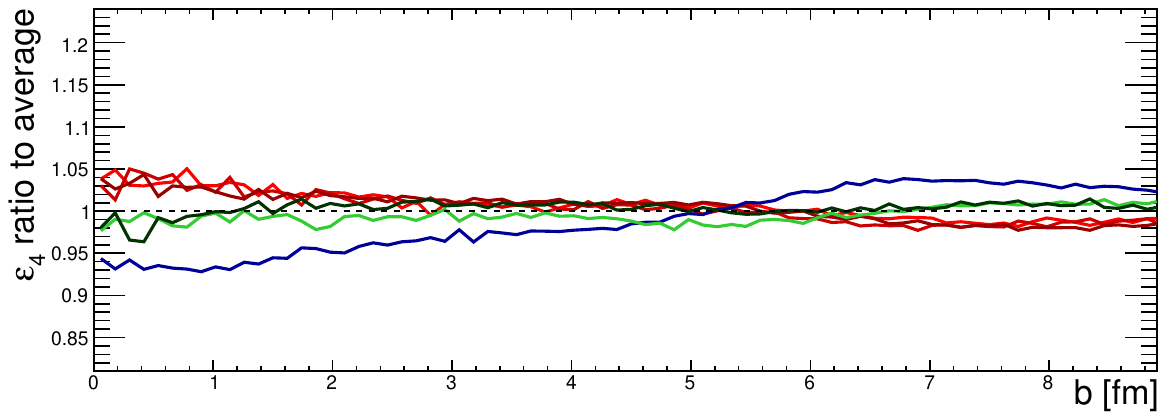}
\caption{Initial-state $\varepsilon_{3}$~(top) and $\varepsilon_{4}$~(bottom) versus impact parameter for \OO~(left column) and \NeNe~(right column) collisions at 5.36~TeV using TGlauberMC with $\omega=0.3$ for the nuclear density profiles discussed in \Sec{sec:nuclearprof}. The smaller bottom panels show the respective ratio to the average over all density profiles.}
\label{fig:extraquantities2}
\end{figure}

By construction, the participant eccentricity approaches unity in the most peripheral collisions when it is calculated directly from the participant positions, which in the limiting case may consist of only two participants. This behavior was first investigated in \Refe{PHOBOS:2007vdf}, which demonstrated for \AuAu\ collisions that applying (Gaussian) smearing to the participant positions primarily affects the participant eccentricity in peripheral events (up to approximately $\Npart \sim 30$). 
For small systems, however, this range effectively covers the entire domain of interest, and smearing therefore has a substantial impact. 
To study whether smearing has an effect on the predicted ratio $\varepsilon_2^{\rm \NeNe}/\varepsilon_2^{\rm \OO}$, one commonly relies on the generalized mean for the nuclear overlap function defined as $\left(\frac{(T_A^p+T_B^p)}{2}\right)^\frac{1}{p}$~\cite{Moreland:2014oya} which for $p=1$ reduces to the arithmetic mean and for $p=0$ to the geometric mean (also known as TRENTO initial conditions).
As shown in \Fig{fig:v2moneyplotsmeared}, in both cases the ratio approaches similar large values towards most central collisions as without Gaussian smearing. 
However, the trend with decreasing centrality differs, in particular for the TRENTO~($p=0$) condition, which increases with decreasing centrality.

\begin{figure}[ht!] 
\includegraphics[width=8cm]{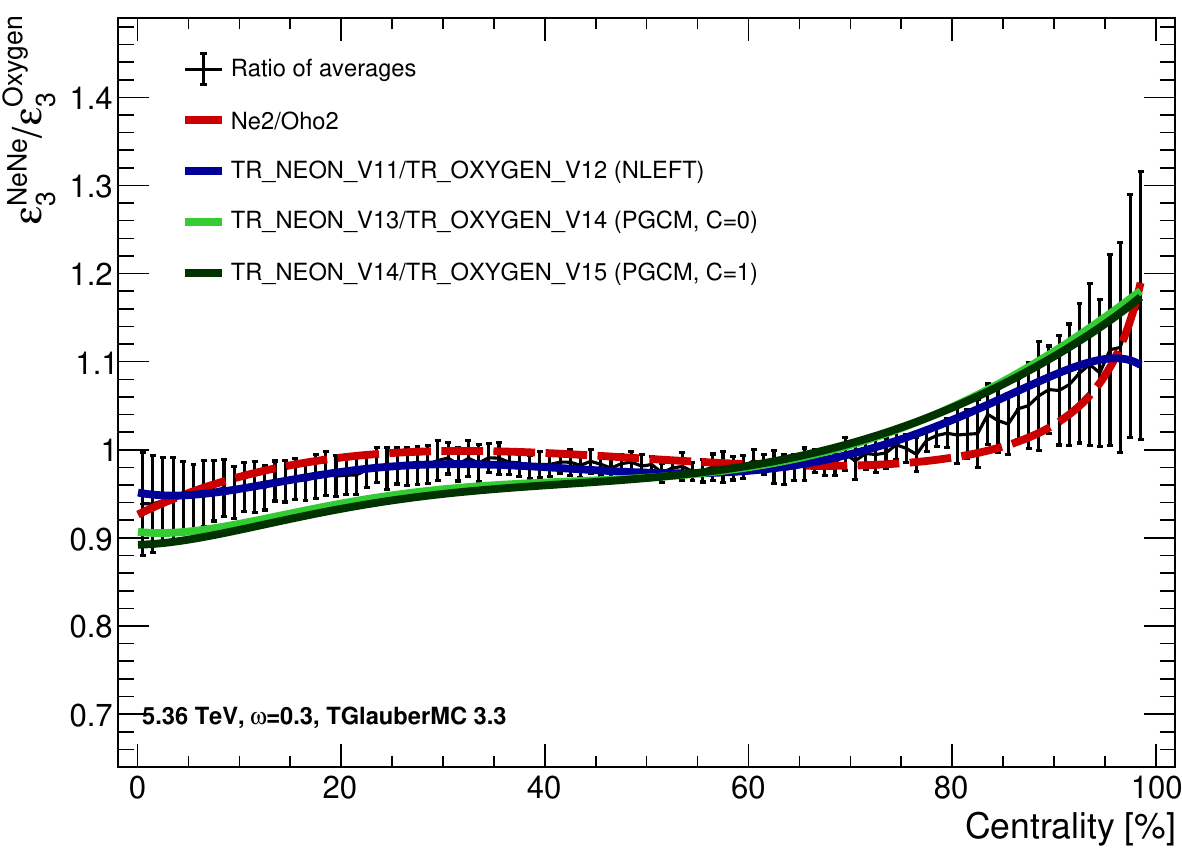}
\includegraphics[width=8cm]{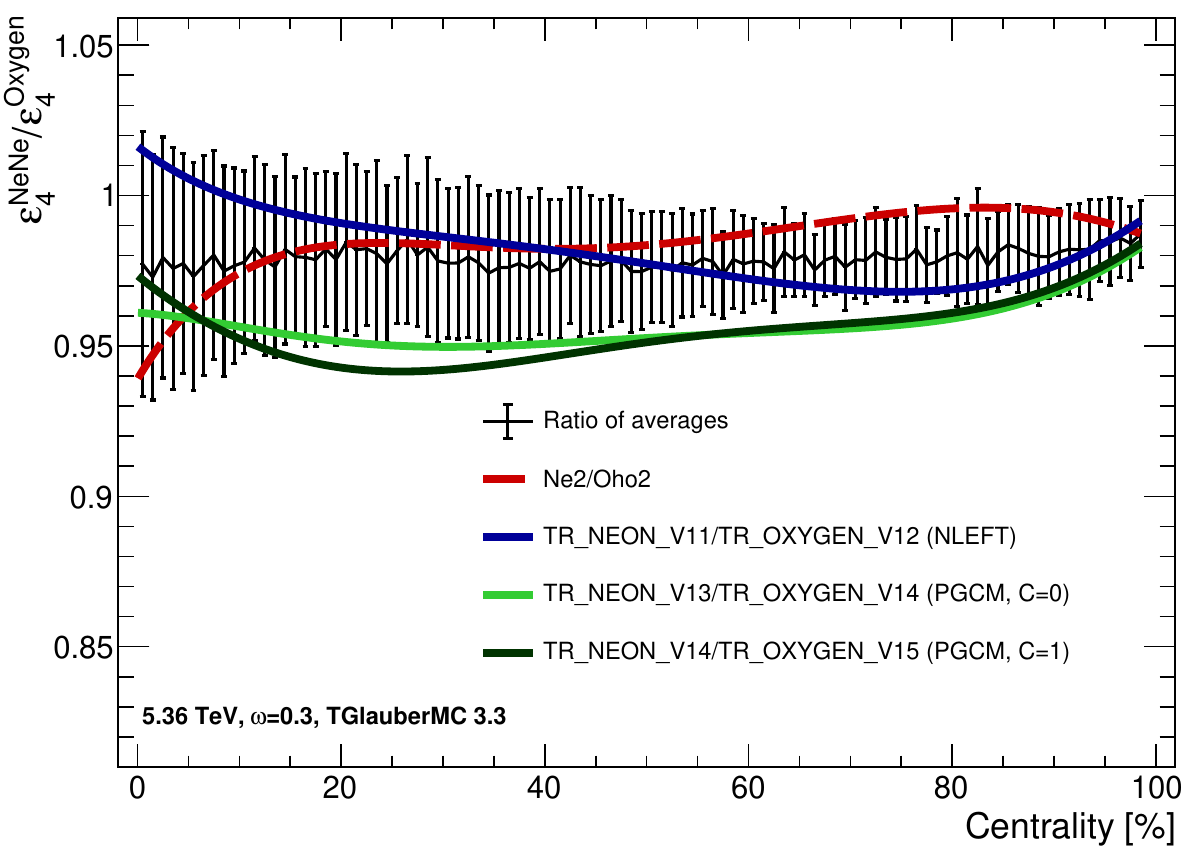}
\caption{Ratio of $\epsilon_3$ (left panel) and $\epsilon_4$ (right panel) versus centrality for \NeNe\ collisions to that for \OO\ collisions at 5.36~TeV using TGlauberMC with $\omega=0.3$ for the ratio of the averages~(standard deviation is shown as uncertainty per point), example ratio using ``Ne2'' and ``Oho2'' parameterizations as well as the ratios obtained using the nuclear densities listed in \Tab{tab:troxygen}.
}
\label{fig:v3moneyplot}
\end{figure}

For completeness, the distributions for $\varepsilon_3$ and $\varepsilon_4$ versus impact parameter for \OO\ and \NeNe\ collisions at 5.36~TeV are shown in \Fig{fig:extraquantities2} using the nuclear density profiles listed in \Tab{tab:troxygen}.
Both exhibit an increasing trend from central to peripheral collisions as known from larger nuclei collision systems.
Variations of up to 10\% relative to the average are observed as a result of the differences between the nuclear density profiles.
\Figure{fig:v3moneyplot} shows the ratios of $\varepsilon_3$ and $\varepsilon_4$ for \NeNe\ collisions to that for \OO\ collisions at 5.36~TeV for the ratio of the averages (including the standard deviation as uncertainty) as well as direct ratios from the nuclear densities listed in \Tab{tab:troxygen}.
The ratio of $\varepsilon_3$ between \NeNe\ and \OO\ exhibits a modest decrease with decreasing centrality, turning into a strong rise for most peripheral collisions.
The ratio of $\varepsilon_4$ between \NeNe\ and \OO\ changes only little with centrality, most notably with a decreasing trend with decreasing centrality for the NLEFT calculation by about 5 percent.

\section{Particle production}
\label{sec:partprod}
The centrality dependence of charged-particle production in heavy-ion collisions was found to scale with the number of participants~\cite{PHOBOS:2010eyu}.
In particular at LHC energies~\cite{Loizides:2011ys}, inclusion of a small contribution from the number of collisions improves the description to account for the increase of multiplicity in more central collisions from mini-jets, as ${\rm d}N/{\rm d}\eta=\alpha \Npart + \beta \Ncoll$.
Typically, ${\rm d}N/{\rm d}\eta$ is normalized per participant pair, $\Npart/2$.

When this formalism is applied across different collision energies and system sizes, employing sub-nucleon degrees of freedom by decomposing each nucleon into three or more effective quark-level interaction centers—removes the need to reference the number of binary nucleon–nucleon collisions, as approximate scaling arises from the generalized number of participant quark centers (see \cite{Loizides:2016djv} and references therein).

\begin{figure}[th!]
\includegraphics[width=8cm]{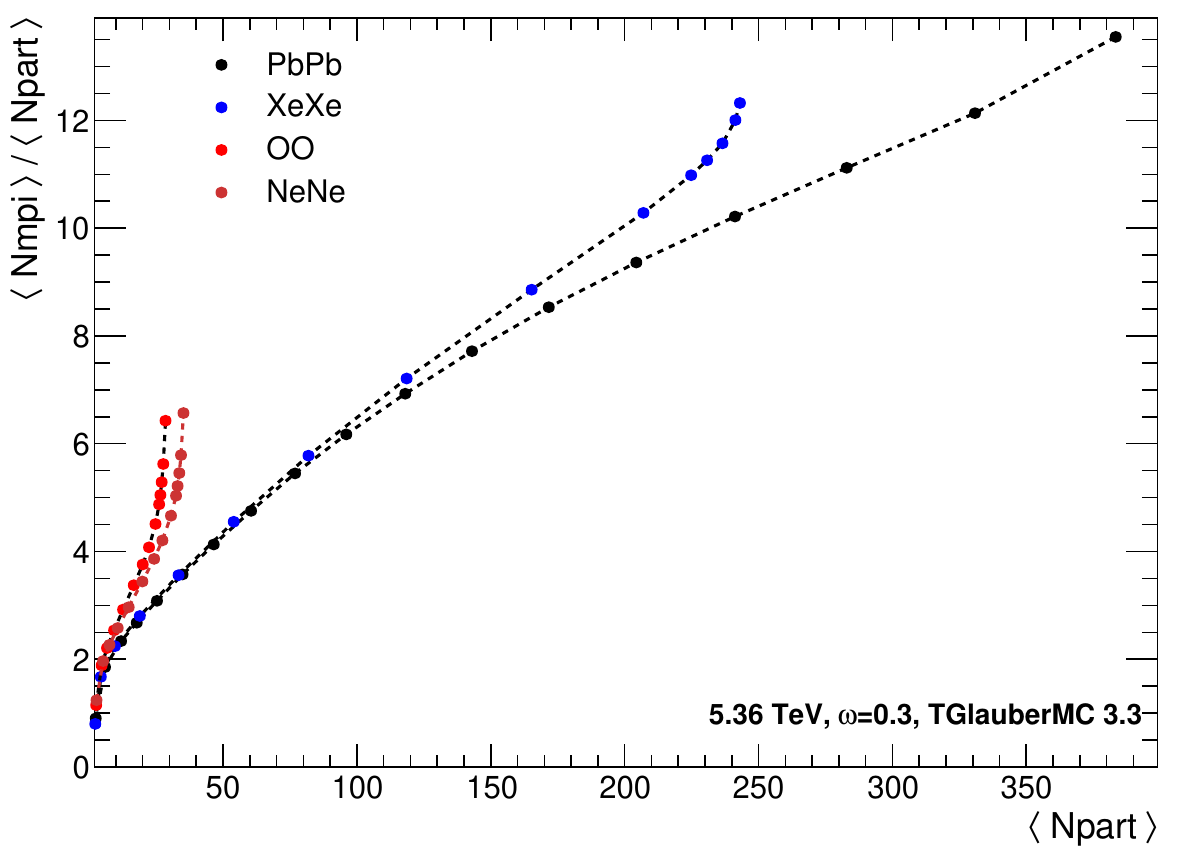}
\includegraphics[width=8cm]{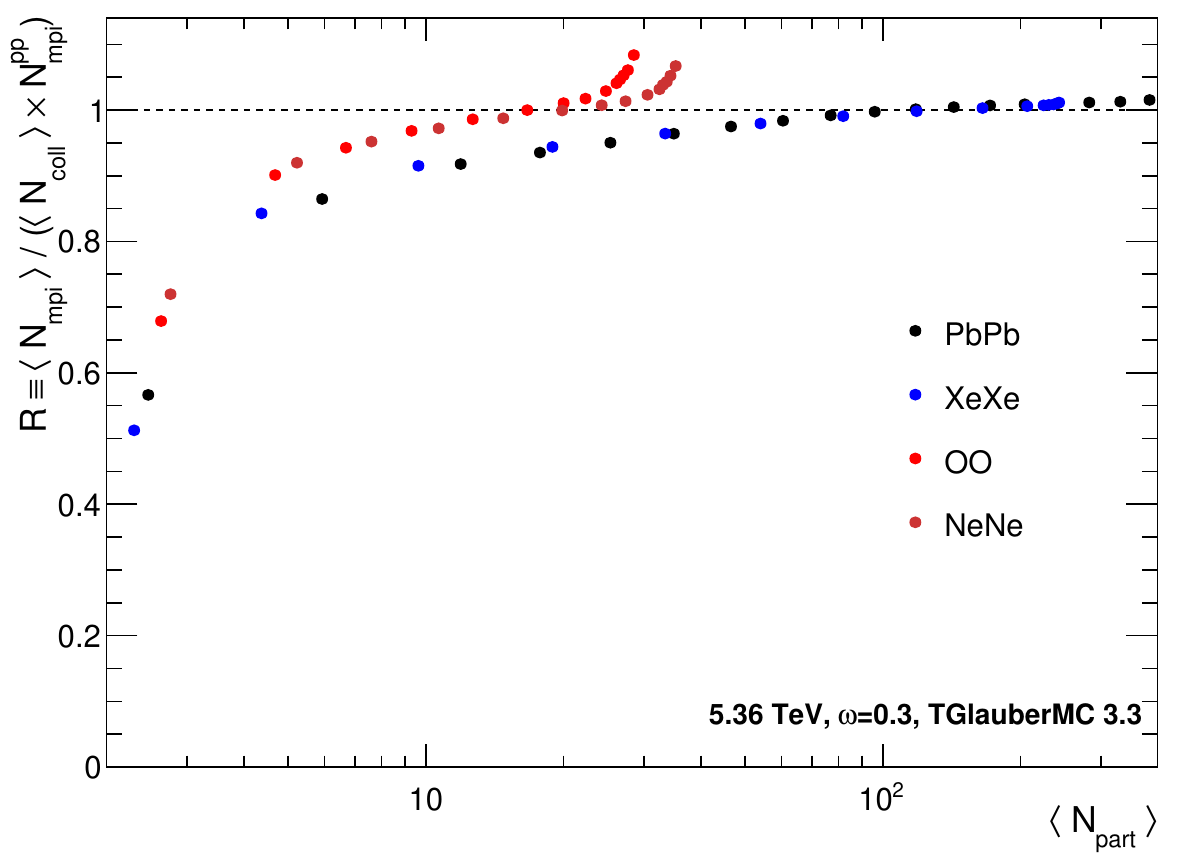}
\caption{Average $\Nmpi/\Npart$~(left panel) and $R\equiv\avg{\Nmpi}/(\avg{\Ncoll}\,\avg{N_{\rm mpi}^{\rm pp}})$~(right panel) versus $\Npart$ for \OO, \NeNe, \XeXe\ and \PbPb\ collisions at 5.36~TeV using TGlauberMC with $\omega=0.3$.}
\label{fig:mpi}
\end{figure}

Here, I take a different approach, and introduce the number of multiple hard interactions as implemented in HIJING~\cite{Wang:1991hta} and applied recently in \cite{Loizides:2017sqq,Jonas:2021xju}.
For a given nucleon--nucleon collision the number of hard scatterings at $\bnn$ is distributed according to a Poisson distribution  with the average given by $\langle \Nmpinn \rangle = \sigma_{\rm NN}^{\rm hard} \, \TNN(\bnn)$, where $\sigma_{\rm NN}^{\rm hard}$ is the energy-dependent pQCD cross-section for $2\to2$ parton scatterings~($\approx190$~mb at 5.36~TeV).
The total number of multiple interactions $\Nmpi$ is then the sum of the multiple interactions over all nucleon--nucleon collision in the event, and on average equal to $\Ncoll$ times the average $\Nmpi$ in \pp\ collisions at the same energy. 
Since the $\Nmpi$ and event multiplicity are strongly correlated~\cite{Loizides:2021ima}, ordering events according to $\Nmpi$ can be used at a proxy for event categorization usually done to group events into centrality intervals.
In this case, the relation on the average between $\Nmpi$ and $\Ncoll$ does no longer need to hold.
The left panel of \Fig{fig:mpi} shows the resulting $\avg{\Nmpi}$ distributions normalized by $\avg{\Npart}$ in centrality-binned intervals versus $\Npart$ for \OO, \XeXe\ and \PbPb\ collisions at 5.36~TeV. 
After a steep rise at very low $\Npart$, the $\Nmpi$ in the different systems rise approximately linearly with $\Npart$, until they eventually increase faster than linear with $\Npart$, most evidently seen for \OO\ and \XeXe\ collisions. 
In the right panel, $R\equiv\avg{\Nmpi}/(\avg{\Ncoll}\,N_{\rm mpi}^{\rm pp})$ versus $\Npart$ is shown, which exhibits the effect of the geometry and event selection bias on the nuclear modification factor, even in absence of any nuclear effect~\cite{Loizides:2017sqq}.
Indeed, the ratio significantly deviates from one in peripheral collisions, where the ratio is smaller up to about 50\%, and while it is enhanced by up to 10\% in the most central \OO\ collisions.
In reality, effects from fragmentation, momentum conservation, play a role, which are expected to result in a $\pt$-dependent selection bias~\cite{Park:2025mbt}.

\begin{figure}[th!]
\includegraphics[width=8cm]{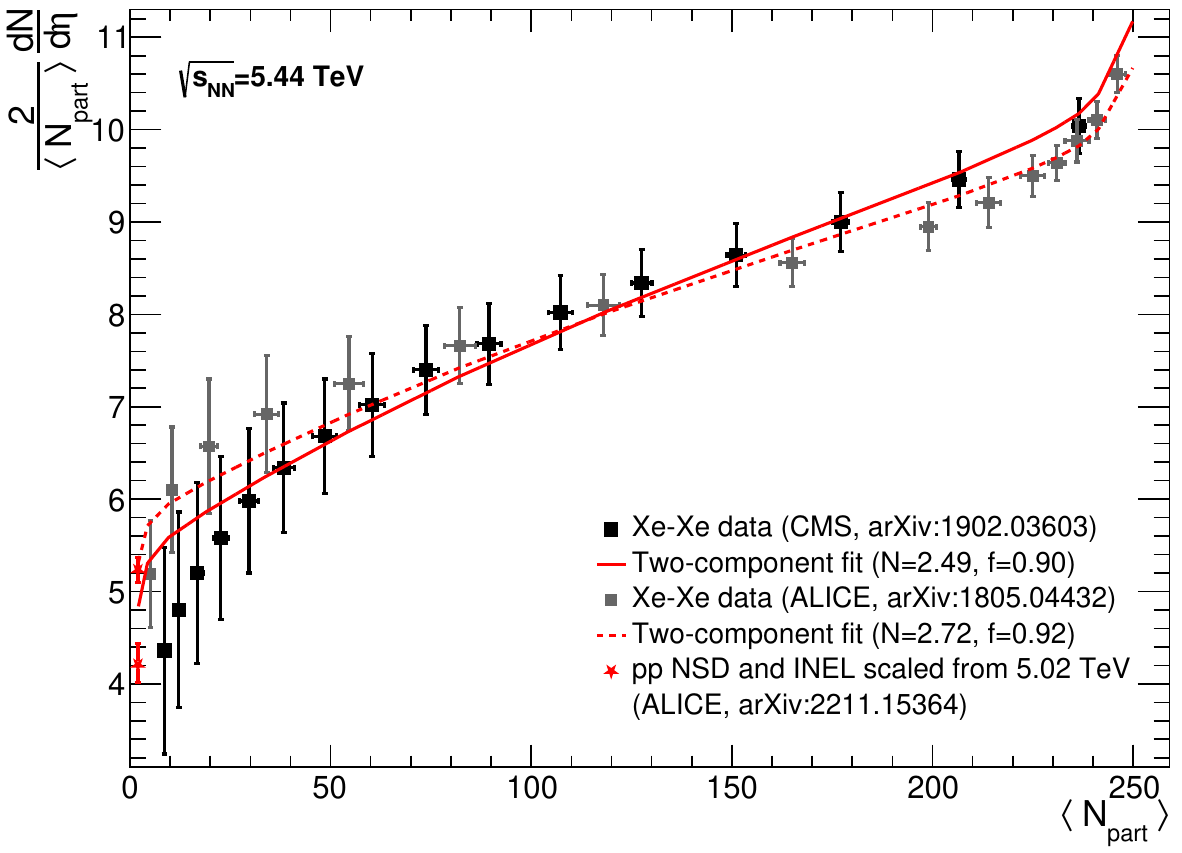}
\includegraphics[width=8cm]{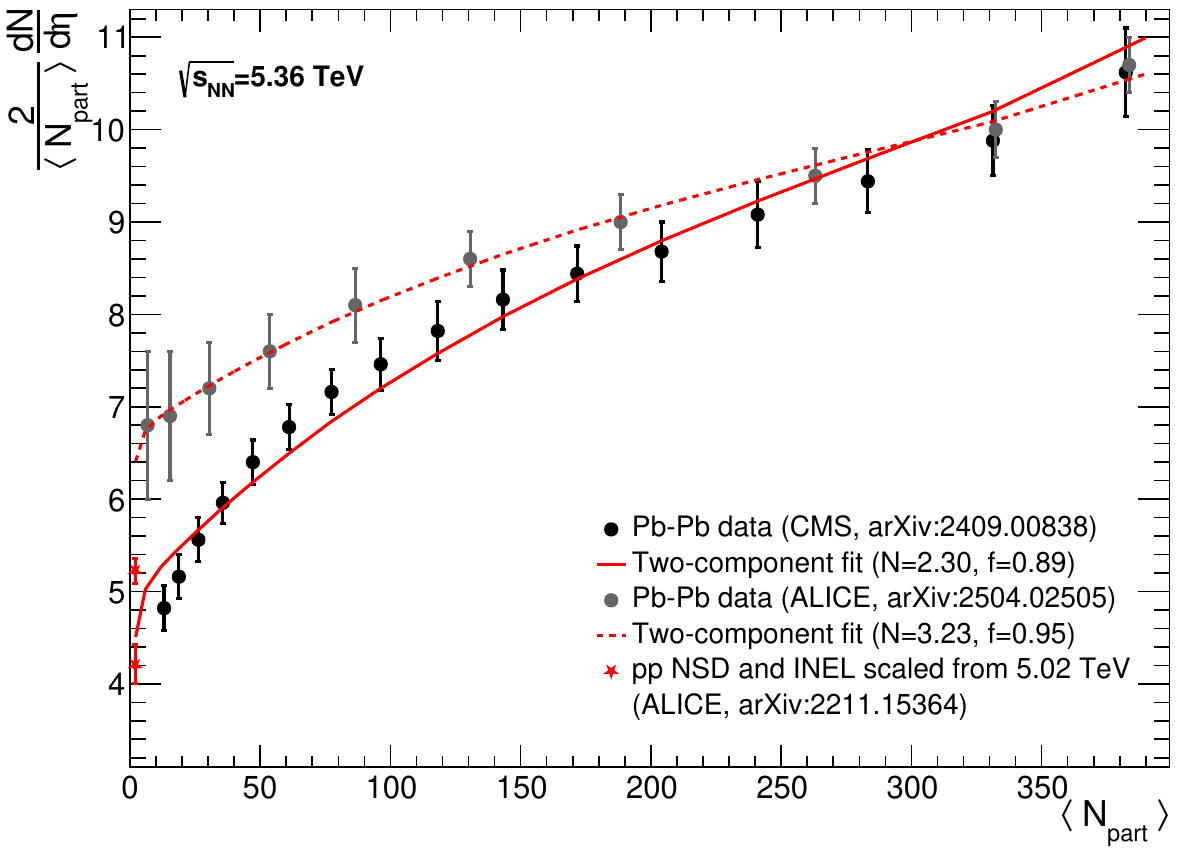}
\caption{Per-participant charged particle pseudo-rapidity density $2/\avg{\Npart}\,{\rm d}N/{\rm d}\eta$ versus $\avg{\Npart}$ at mid-rapidity in \XeXe\ collisions at 5.44~TeV~(left) and in \PbPb\ collisions at 5.36~TeV~(right) obtained by ALICE and CMS~\cite{ALICE:2018cpu,CMS:2019gzk,CMS:2024ykx,ALICE:2025cjn}, as well as two component fits to each dataset~(see text).
Data~\cite{ALICE:2022kol} for \pp\ collisions normalized to non-single diffractive and inelastic yields are also shown.
}
\label{fig:dndeta}
\end{figure}

Per-participant charged particle pseudo-rapidity density $2/\avg{\Npart}\,{\rm d}N/{\rm d}\eta$ versus $\avg{\Npart}$ at mid-rapidity in \XeXe\ collisions at 5.44~TeV and in \PbPb\ collisions at 5.36~TeV obtained by ALICE and CMS~\cite{ALICE:2018cpu,CMS:2019gzk,CMS:2024ykx,ALICE:2025cjn} are shown in \Fig{fig:dndeta}.
In case of \XeXe\ collisions the ALICE and CMS date are not consistent for peripheral collisions.
Data for \pp\ collisions normalized to non-single diffractive and inelastic yields are also shown~(scaled to 5.36~TeV by less than one percent as given by respective power laws as a function of $\sqrt{s}$)~\cite{ALICE:2022kol}.

\begin{figure}[th!] 
\includegraphics[width=10.4cm]{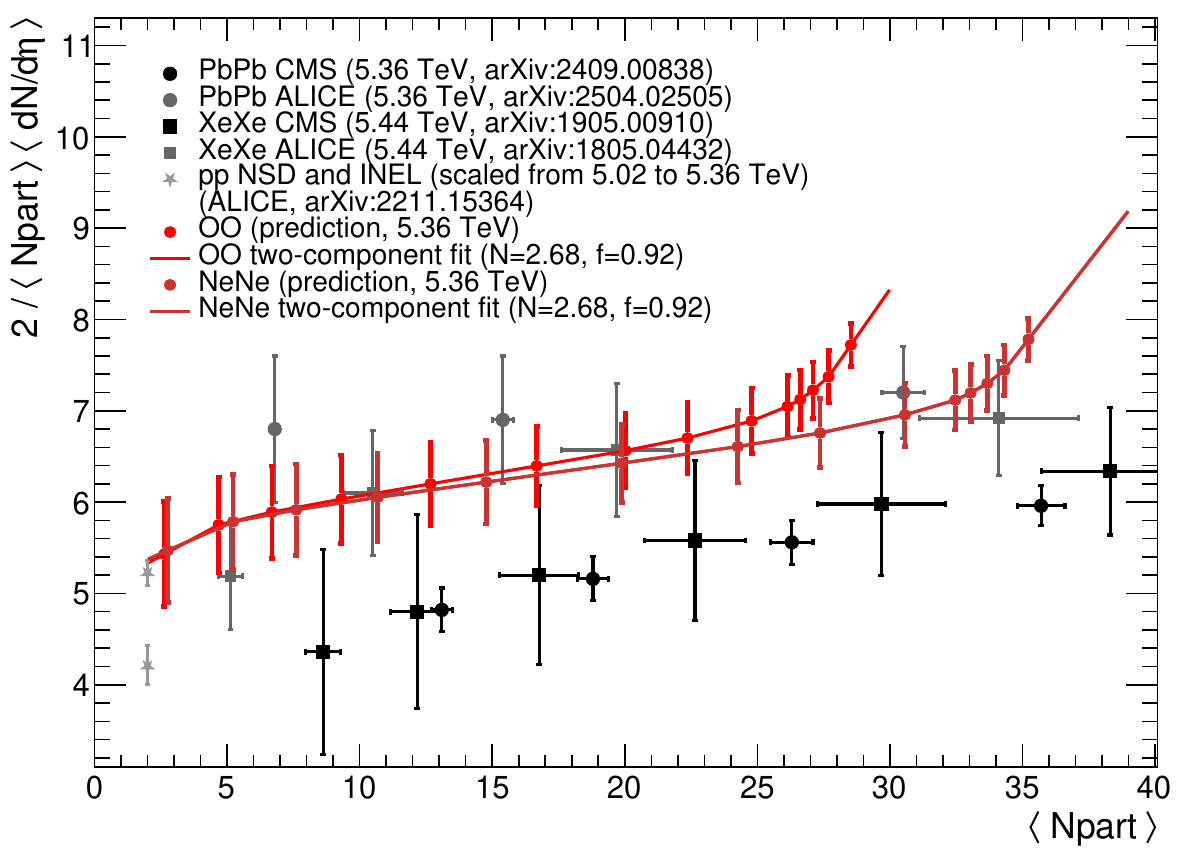}
\caption{Per-participant charged particle pseudo-rapidity density $2/\avg{\Npart}\,{\rm d}N/{\rm d}\eta$ versus $\avg{\Npart}$ at mid-rapidity in \XeXe\ collisions at 5.44~TeV and in \PbPb\ collisions at 5.36~TeV obtained by ALICE and CMS~\cite{ALICE:2018cpu,CMS:2019gzk,CMS:2024ykx,ALICE:2025cjn}, as well as data for \pp\ collisions normalized to non-single diffractive and inelastic yields~\cite{ALICE:2022kol}.
The figures zooms into the low $\Npart$ region to show highlight predictions for \OO\ and \NeNe\ collisions at 5.36~ TeV obtained by average of the fits obtained in \Fig{fig:dndeta} together with the respective two-component model fits~(see text).
}
\label{fig:dndeta2}
\end{figure}

Despite the apparent opposite trend in the \PbPb\ case the dataset are consistent considering the large uncertainties in peripheral collisions.
Since the data are essentially at the same collision energy as that for the planned \OO\ collisions, they can be used to predict the ${\rm d}N/{\rm d}\eta$ for \OO\ collisions, by fitting a two-component model $2N(f+(1-f)\,{\avg{\Nmpi}}/{\avg{\Npart}})$.
The fits are found to describe the data reasonably well, however with different values for $N$ between \XeXe\ and \PbPb\ collisions, in particular for ALICE. 
Hence, for the predicted result, the central value between the ratios is used with half of the spread as uncertainty assigned to essentially cover all cases.
The same procedure was also applied to give the expected mid-rapidity ${\rm d}N/{\rm d}\eta$ for \NeNe\ collisions.
The expectations for \OO\ and \NeNe\ collisions at 5.36~TeV are shown in \Fig{fig:dndeta2} together with the existing data.

\begin{table}[th!]
\centering
\caption{Centrality dependence of $N_{\rm part}$, $N_{\rm coll}$, $N_{\rm mpi}$, $R\equiv\avg{\Nmpi}/(\avg{\Ncoll}\,\avg{N_{\rm mpi}^{\rm pp}}$ and $\frac{2}{\Npart}{\rm d}N/{\rm d}\eta/\Npart$ in \OO\ and \NeNe\ collisions at 5.36~TeV. The Glauber values are given without uncertainties.}
\begin{tabular}{|c|ccccc|ccccc|}
\hline
& \multicolumn{5}{c|}{\OO} & \multicolumn{5}{c|}{\NeNe} \\
Centrality & $\langle N_{\rm part}\rangle$ & $\langle N_{\rm coll}\rangle$ & $\langle N_{\rm mpi}\rangle$ & $R$ & $\frac{2}{\Npart}{\rm d}N/{\rm d}\eta$ & $\langle N_{\rm part}\rangle$ & $\langle N_{\rm coll}\rangle$ & $\langle N_{\rm mpi}\rangle$ & $R$ & $\frac{2}{\Npart}{\rm d}N/{\rm d}\eta$ \\
\hline
0--1\% & 28.5 & 57.7 & 183.3 & 1.084 & 7.72$\pm$0.24 & 35.2 & 73.9 & 231.4 & 1.067 & 7.78$\pm$0.23 \\
1--2\% & 27.7 & 50.1 & 155.7 & 1.061 & 7.37$\pm$0.29 & 34.3 & 64.4 & 198.6 & 1.052 & 7.44$\pm$0.28 \\
2--3\% & 27.1 & 46.4 & 143.2 & 1.053 & 7.23$\pm$0.31 & 33.7 & 60.1 & 183.7 & 1.043 & 7.30$\pm$0.30 \\
3--4\% & 26.6 & 43.8 & 134.4 & 1.047 & 7.12$\pm$0.32 & 33.0 & 56.6 & 172.3 & 1.038 & 7.20$\pm$0.31 \\
4--5\% & 26.1 & 41.8 & 127.4 & 1.041 & 7.05$\pm$0.34 & 32.5 & 54.0 & 163.4 & 1.032 & 7.12$\pm$0.33 \\\hline
0--5\% & 27.2 & 48.0 & 149.1 & 1.059 & 7.30$\pm$0.30 & 33.8 & 61.9 & 190.3 & 1.05 &  7.37$\pm$0.29   \\
5--10\% & 24.8 & 37.0 & 111.8 & 1.029 & 6.89$\pm$0.36 & 30.6 & 47.5 & 142.5 & 1.023 & 6.96$\pm$0.35 \\
10--15\% & 22.4 & 30.6 & 91.2 & 1.017 & 6.70$\pm$0.39 & 27.4 & 38.7 & 115.1 & 1.013 & 6.76$\pm$0.38 \\
10--20\% & 20.0 & 25.4 & 75.3 & 1.011 & 6.56$\pm$0.41 & 24.3 & 31.7 & 93.7 & 1.007 & 6.61$\pm$0.40 \\
20--30\% & 16.7 & 19.2 & 56.3 & 1.000 & 6.40$\pm$0.43 & 19.9 & 23.4 & 68.4 & 0.999 & 6.43$\pm$0.43 \\
30--40\% & 12.7 & 12.8 & 37.0 & 0.986 & 6.20$\pm$0.46 & 14.8 & 15.1 & 43.8 & 0.987 & 6.22$\pm$0.46 \\
40--50\% & 9.3 & 8.3 & 23.6 & 0.968 & 6.03$\pm$0.48 & 10.7 & 9.7 & 27.5 & 0.972 & 6.05$\pm$0.48 \\
50--60\% & 6.7 & 5.3 & 14.8 & 0.943 & 5.89$\pm$0.51 & 7.6 & 6.2 & 17.2 & 0.952 & 5.92$\pm$0.50 \\
60--70\% & 4.7 & 3.3 & 8.8 & 0.901 & 5.75$\pm$0.53 & 5.2 & 3.8 & 10.3 & 0.920 & 5.79$\pm$0.52 \\
70--100\% & 2.6 & 1.5 & 3.0 & 0.679 & 5.43$\pm$0.57 & 2.8 & 1.6 & 3.4 & 0.720 & 5.47$\pm$0.57 \\\hline
0--100\% & 10.8 &12.8 &37.6 & 1 & $6.02\pm0.21$ & 12.7 & 15.7 & 46.0 & 1 & $6.06\pm0.21$ \\
\hline
\end{tabular}
\label{tab:values}
\end{table}

A table with all relevant Glauber related values and projected multiplicities is given in \Tab{tab:values}.
The Glauber values are given without uncertainties.
However, it should be noted that the associated uncertainties usually estimated from data require careful evaluation, and are expected to be non-negligible.

\section{Summary}
\label{sec:summary}
An updated version~(\version) of the TGlauberMC model is presented, which includes several different probability distributions for the nucleon--nucleon interaction probability~(\Fig{fig:oproftglaubermc}), and a huge variety of nuclear density profiles~(\Tab{tab:gloxygen} and \ref{tab:troxygen}), in particular for oxygen and neon~(\Fig{fig:oprofdata} and \ref{fig:oprofdata2}).
The model is used to predict properties of \OO, \NeNe, as well as \pPb\ and \PbPb\ collisions at 5.36~TeV, and \pO\ collisions at 9.62~TeV.
The expected geometrical cross sections are averaged over all provided profiles~(\Fig{fig:oxnecs}), and considered to include a contribution from the transverse nucleon--nucleon overlap~(\Fig{fig:ox_om_cs} and ~\ref{fig:pb_om_cs}), and presented in \Tab{tab:xsres}.
For central collisions, the distributions of $\Ncoll$ and $\Npart$ exhibit a spread of 5--10\% and about 2\%, respectively~(\Fig{fig:quantities}), and up to 20\% for the participant eccentricity~(\Fig{fig:quantities2}).
The ratio of $\varepsilon_{2}$ in \NeNe\ to \OO\ is found to reach up to $1.15\pm0.05$~(\Fig{fig:v2moneyplot}), rather independently whether Gaussian smearing~(\Fig{fig:v2moneyplotsmeared}) using generalized TRENTO conditions is used or not.
Rising trends for $\varepsilon_{3}$ and $\varepsilon_{4}$ with decreasing centrality~(\Fig{fig:extraquantities2}), while first decreasing then rising trends with decreasing centrality of the respective ratios between \NeNe\ and \OO\ collisions~(\Fig{fig:v3moneyplot}) are found.
Using $\Nmpi$ instead of $\Ncoll$~(\Fig{fig:mpi}) allows to predict general trends of the \OO\ data~(\Fig{fig:dndeta})
and use existing data at 5.36 TeV to predict the multiplicity in \OO\ and \NeNe\ collisions~(\Fig{fig:dndeta2}).
The source code for TGlauberMC~(\version) is available at HepForge~\cite{glaucode}, and the author welcomes suggestions how to improve it.

During the peer-review of the manuscript results from \OO\ collisions at $\snn=200$~GeV, taken in June, 2021 at RHIC\cite{STAR:2025ivi}, and from \OO\ and \NeNe\ collisions at $5.36$~TeV at LHC~\cite{ALICE:2025luc,ATLAS:2025nnt,CMS:2025tga,ATLAS:2025ooe,CMS:2025bta,2969907} were submitted.
The availability of the \OO\ data at both RHIC and LHC, opens an important opportunity to explore both particle production and collective flow across a wide energy range — and TGlauberMC~(\version) should prove a valuable tool for interpreting those measurements.

\begin{acknowledgments}
The author would like to thank G.~Nijs and W.~van der Schee for fruitful discussions related to nuclei configurations provided by Trajectum,
G.~Giacalone for clarification related to the 3pF fit of the NNLOsat oxygen profile, 
V.~Soma for providing the exact data points of the NNLOsat calculation~\cite{Soma:2019bso},
B.~Kardan for contributing changes in TGlauberMC~v3.3 useful for Hades/CBM collision energies, 
as well as F.~Jonas and N.~Strangmann for taking the time to read and comment on a draft of this article.
The author acknowledges financial support by the U.S.\ Department of Energy, Office of Science, Office of Nuclear Physics, under contract number DE-SC0005131.
\end{acknowledgments}
\bibliographystyle{utphys}
\bibliography{biblio}
\appendix 
\label{sec:App}
\section{TGlauberMC v3.3 user's guide}
\label{app:guide}
The main changes related to the previous version~(3.2) is the interface to any nucleus configuration provided by Trajectum~\cite{Nijs:2021clz} via
the new TrNucGen~\cite{trnucgen} library~(see \App{sec:trnucgen}), the support to read nucleus configurations from text files, and a larger list pre-defined nucleus configurations.
Further, the possibility to use between overlap functions using TRENTO, HIJING and PYTHIA parameterizations, in addition to the $\Gamma$-distribution that already was present since v3.0. 
As default, the code continues to use the hard-sphere approximation~($\omega=0$, see \Sec{sec:transoverlap}).
Following the discussion in \Sec{sec:partprod}, the total number of multiple interactions are computed from the nucleon--nucleon overlap function.
The $\sigmaNN$ parameterization for RHIC and LHC energies is provided, as well as dedicated parameterizations~\cite{Bystricky:1987yq} for neutron--neutron~(proton--proton) and neutron--proton interactions relevant at Hades/CBM energies.

The user's guide has been incrementally updated from the first release of the code in 2008~\cite{Alver:2008aq}, the improved version in 2014~\cite{Loizides:2014vua}, and the latest in 2017~\cite{Loizides:2017ack}.
As before, the source code, which relies on the ROOT~\cite{Brun:1997pa} framework (version 6.x), can be obtained at the TGlauberMC page on HepForge~\cite{glaucode}.
For version 3.3, it also provides a ``\href{https://tglaubermc.web.cern.ch/html/index.html}{doxygen}'' documentation, which provides many details.
In the following, I only describe the improved functionality relative to the last update~\cite{Loizides:2017ack}.

All functionality is implemented in the macro {\tt runglauber\_vX.Y.C}, where version ``X.Y==3.3'' described here.
It is best to use the provided ``rootlogon.C'' macro in the same directory which will setup the code automatically. 
Unlike before, additional input files for nucleus configurations are now located in the subdirectory ``dat''.
Three classes, {\tt TGlauNucleon}, {\tt TGlauNucleus} and {\tt TGlauberMC} and five functions~(macros) {\tt runAndSaveNtuple()}, {\tt runAndSaveNucleons()}, {\tt runAndSmearNtuple()}, {\tt runAndOutputLemonTree()} and {\tt runAndCalcDens()} are defined in the provided macro as in earlier versions.

As before, the most convenient starting point is the provided {\tt runAndSaveNtuple()} which will generate the requested number of \MCG\ events for given nucleus types, $\sigmaNN$ as well as most notably the $\dmin$ and $\omega$ values, storing the event-by-event computed quantities in a ROOT tree. 
The only difference to earlier versions is that if the provided value for $\sigmaNN$ is negative, it will be interpreted as beam energy to compute
$\sigmaNN$ and $\sigma_{\rm NN}^{\rm hard}$.
The updated list of supported nuclei can be found in the {\tt TGlauNucleus::Lookup} function in the code.
Nuclei names that start with ``TR'' are taken from TrNucGen, see \App{sec:trnucgen}.
For this to work, the TrNucGen library must be available in a subdirectory called ``trnucgen''~(or a symbolic link must be present that points to it).
Nuclei names that start with ``input'' are read from a text file with the following structure:
\begin{verbatim}
# Nucleus generator version: 1.0.0
# Nucleus type: TR_CARBON_V1
# Nucleus id: 7
# Nucleus name: C1
# Nucleus description: carbon12(v1)
# Number of nucleons: 12
# Number of protons: 6
# Number of events: 10000
# Per line all nuclei, in the format x1, y1, z1, c1, x2, y2, z2, c2, ... 
\end{verbatim}
where the position ($x$, $y$, $z$) and charge $c$ are listed per line for a nucleus configuration.
The format of the file is also explained in the example provided with TrNucGen~\cite{trnucgen}).

The following set of functions controls additional behavior of the {\tt TGlauNucleus} class:
{\tt GetDens} returns the density normalized to $Z$ or to $A$.
{\tt GetSqrtMeanR2()} returns the $r_{\rm rms}$ computed from the density distribution.
{\tt CalcMinDist()} calculates the smallest difference between any two nucleons in the nucleus.
{\tt CalcRmsRadius()} calculates the $r_{\rm rms}$ computed from the nucleon positions assigned to the nucleus.
{\tt GetWeight{}} and {\tt SetWeight()} returns or sets the weight for the particular nucleon configuration (only used by nuclei constructed with TrNucGen).
{\tt SetBeta(Double\_t b2, Double\_t b3, Double\_t b4, Double\_t g)} allows to set an almost deformation parameters $\beta_2$, $\beta_3$, $\beta_4$ and $\gamma$ that together with ``box'' Monte Carlo sampling allows the generation for deformed nuclei.  

The following set of member functions controls additional behavior of the {\tt TGlauberMC} class:  
{\tt SetOmega($w$)} to provide the nucleon--nucleon profile to use. The input value can be $0<w<2$ for the $\Gamma$-distribution, $7$ for HIJING, $8$ for PYTHIA and $9<w<11$, which would translate into $0<\delta<2$ for TRENTO.
{\tt SetSigmaHard} to set the hard scattering cross section (in case you do not want to rely on the precomputed value.
{\tt SetShadowing} to control whether or not shadowing corrections are computed.

In addition to quantities described the previous user guides, the following quantities are stored in the ROOT tree:
\begin{itemize}
 \item {\tt Nmpi}: Number of MPI (HIJING model)
 \item {\tt NpartAn}: Number of wounded (participating) neutrons in Nucleus A
 \item {\tt NpartBn}: Number of wounded (participating) neutrons in Nucleus B
 \item {\tt Npart0n}: Number of singly-wounded (participating) neutrons
 \item {\tt SpecA}: Spectator neutrons in nucleus A
 \item {\tt SpecB}: Spectator neutrons in nucleus B
 \item {\tt Weight}: Weight of event (needed for e-by-e weighting)
\end{itemize}

Additional global functions that are provided are 
\begin{itemize}
\item {\tt getNNProf()} and {\tt getNNProfDist} for the $\Gamma$-based collision probability
\item {\tt getNNHijing()} and {\tt getNNHijingDist()} for the HIJING collision probability
\item {\tt getNNPythia()} and {\tt getNNPythiaDist()} for the PYTHIA collision probability
\item {\tt getNNTrento()} and {\tt getNNTrentoDist()} for the TRENTO collision probability
\item {\tt getSigmaNNvsEnergy()} to get $\sigmaNN$ as function of collision energy
\item {\tt getSigmaNN()} to return the $\sigmaNN$ for a given energy
\item {\tt getSigmaNPvsEnergy\_Bystricky()} and {\tt getSigmaPPvsEnergy\_Bystricky()} to get $\sigma_{\rm NN}$ and $\sigma_{\rm NP}$ as function of energy (useful mainly below 10 GeV)
\item {\tt getSigmaNP\_Bystricky()} and {\tt getSigmaPP\_Bystricky()} to get $\sigma_{\rm NN}$ and $\sigma_{\rm NP}$ (useful mainly below 10 GeV)
\item {\tt getSigmaHardvsEnergy()} returns $\sigma_{\rm NN}^{\rm hard}$ as a function of energy
\item {\tt getSigmaHard()} returns  $\sigma_{\rm NN}^{\rm hard}$ at the given energy
\end{itemize}
These also exists as corresponding static functions of the {\tt TGlauberMC} class.
Additional static functions are {\tt ReadNtuple} to read the produced Ntuple, {\tt GetCentralityDist} to produce a centrality distribution based on impact parameter and {\tt AddCentralityBranch} to add the centrality estimator to the produced tree.

The {\tt runAndSmearNtuple()} function has been improved to include generalized overlap function defined as $\left(\frac{(T_A^p+T_B^p)}{2}\right)^\frac{1}{p}$~\cite{Moreland:2014oya}, which for $p=1$ reduces to the arithmetic mean and for $p=0$ to the geometric mean (also known as TRENTO initial conditions).
The function produces a sample of $n$ events with a generalized reduced-thickness (or $p$-parameter) formulation of the overlap.  The parameter $p$ controls the functional form of the participant thickness; unless stated otherwise the default value $p=1$ is used, while $p=0$ corresponds to the TRENTO prescription.
The nuclei involved in the collision are specified by the strings \texttt{sysA} and \texttt{sysB}, which are interpreted through the
\texttt{TGlauNucleus::Lookup} tables.  
The inelastic nucleon--nucleon cross section is set by $\sigma_{NN}$ (in units of mb), for which a default value of $68\,\mathrm{mb}$ is used.  
A minimum nucleon–nucleon separation $d_{\min}$ is enforced during nucleon placement, with
$d_{\min}=0.4\,\mathrm{fm}$ in our baseline configuration.
The model allows for a tunable nucleon profile characterized by the parameter $\omega$.  
A value $\omega=0$ corresponds to a hard-sphere profile, $0\le\omega\le2$ yields a Gamma-distributed profile, and $\omega=7$ and $8$ reproduce the effective profiles used in HIJING and PYTHIA, respectively.  
Larger values ($\omega=9$--$11$) correspond to the generalized TRENTO implementations.  
Additional profile-shape parameters include $w$, for which the TRENTO-motivated value $w=0.5$ is used, as well as $k$, the shape parameter of the Gamma distribution, with default value $k=1.4$.
Events are sampled within an impact-parameter range $b_{\min} \le b \le b_{\max}$, where $b_{\min}=0$ and $b_{\max}=20\,\mathrm{fm}$ are used unless otherwise stated.  
The resulting events are written to an output ntuple specified by \texttt{fname}; if no
file name is supplied, a default name is automatically generated.

\section{TrNucGen}
\label{sec:trnucgen}
The Trajectum Nucleus Generator (TrNucGen) software library~\cite{trnucgen} provides all nuclei configurations available in Trajectum~v2.1~\cite{Nijs:2021clz}.
It was written in particular to use all the configurations for Oxygen-16 and Neon-20 nuclei~\cite{Giacalone:2024luz} explored in this paper.
The library is defines a convenient interface to the relevant code extracted from Trajectum, avoiding having to compile the latter with all the required dependencies.
The full list of available oxygen and neon profiles from Trajectum is given in \Tab{tab:troxygenall}.
The list of all nuclei configurations can be found in the doxygen documentation of the code.
The installation instructions as well as the source code including a standalone example are publicly available~\cite{trnucgen}.
The library can be also very useful for any heavy-ion model or simulation that would like to explore nuclei configurations beyond those based on common parameterizations~(i.e.\ 3pF). 

\begin{table}[th!]
\centering
\caption{Full list of oxygen-16 and neon-20 NLEFT and PGCM calculations taken from Trajectum~\cite{Giacalone:2024luz}.}
\begin{tabular}{ll}
\hline
Profile & Description \\
\hline
TR\_OXYGEN\_V1 & Variational Monte Carlo version \cite{Lim:2018huo} (same as ``O'' in TGlauberMC v2.7 and higher) \\
TR\_OXYGEN\_V2 & NLEFT pinhole, $\pm$ weights \\  
TR\_OXYGEN\_V3 & 3pF from \cite{Summerfield:2021oex} (same as ``Opar'' in TGlauberMC v2.7 and higher) \\
TR\_OXYGEN\_V4 & 3pF fit to NNLOSat \cite{{Soma:2019bso}} (same as ``Opar2'' in TGlauberMC v3.3) \\
TR\_OXYGEN\_V5 & Ab-initio unprojected PGCM. \\
TR\_OXYGEN\_V6 & NLEFT version with only positive weights \\
TR\_OXYGEN\_V7 & Ab-initio unprojected PGCM, enforcing clustering \\
TR\_OXYGEN\_V8 & Ab-initio unprojected PGCM, chopping off artifact regions \\
TR\_OXYGEN\_V9 & Ab-initio unprojected PGCM, enforcing clustering and chopping off artifact regions.\\
TR\_OXYGEN\_V10 & PGCM, constraints imposed on unprojected states, no explicit clustering \\
TR\_OXYGEN\_V11 & PGCM, constraints imposed on unprojected states, with explicit clustering \\
TR\_OXYGEN\_V12 & NLEFT pinhole, $\pm$ weights, with periodicity ambiguities resolved \\
TR\_OXYGEN\_V13 & NLEFT version with only positive weights, with periodicity ambiguities resolved \\
TR\_OXYGEN\_V14 & PGCM, constraints imposed on projected states, no explicit clustering \\
TR\_OXYGEN\_V15\color{white}aaa\color{black} & PGCM, constraints imposed on projected states, with explicit clustering \\
\hline 
TR\_NEON\_V1 & Ab-initio unprojected PGCM \\
TR\_NEON\_V2 & 3pF plus deformation (same as ``NeTr2'' in TGlauberMC v3.3\\
TR\_NEON\_V3 & Fit to TR\_NEON\_V1 (same as ``NeTr3'' in TGlauberMC v3.3\\
TR\_NEON\_V4 & NLEFT pinhole, $\pm$ weights \\
TR\_NEON\_V5 & NLEFT version with only positive weights\\
TR\_NEON\_V6 & Ab-initio unprojected PGCM, enforcing clustering\\
TR\_NEON\_V7 & Ab-initio unprojected PGCM, chopping off artifact regions.\\
TR\_NEON\_V8 & Ab-initio unprojected PGCM, enforcing clustering and chopping off artifact regions\\
TR\_NEON\_V9  & PGCM, constraints imposed on unprojected states, no explicit clustering \\
TR\_NEON\_V10 & PGCM, constraints imposed on unprojected states, with explicit clustering \\
TR\_NEON\_V11 & NLEFT pinhole, $\pm$ weights, with periodicity ambiguities resolved \\
TR\_NEON\_V12 & NLEFT with only positive weigths, with periodicity ambiguities resolved\\
TR\_NEON\_V13 & PGCM, constraints imposed on projected states, no explicit clustering \\
TR\_NEON\_V14 & PGCM, constraints imposed on projected states, with explicit clustering \\
\hline
\end{tabular}
\label{tab:troxygenall}
\end{table}
\clearpage
\ifarxiv
\section{Additional figures}
\label{sec:addfigs}
This section shows additional figures only available in the arXiv version of this text, and mainly for completeness.

\begin{figure}[thb!]
\includegraphics[width=8cm]{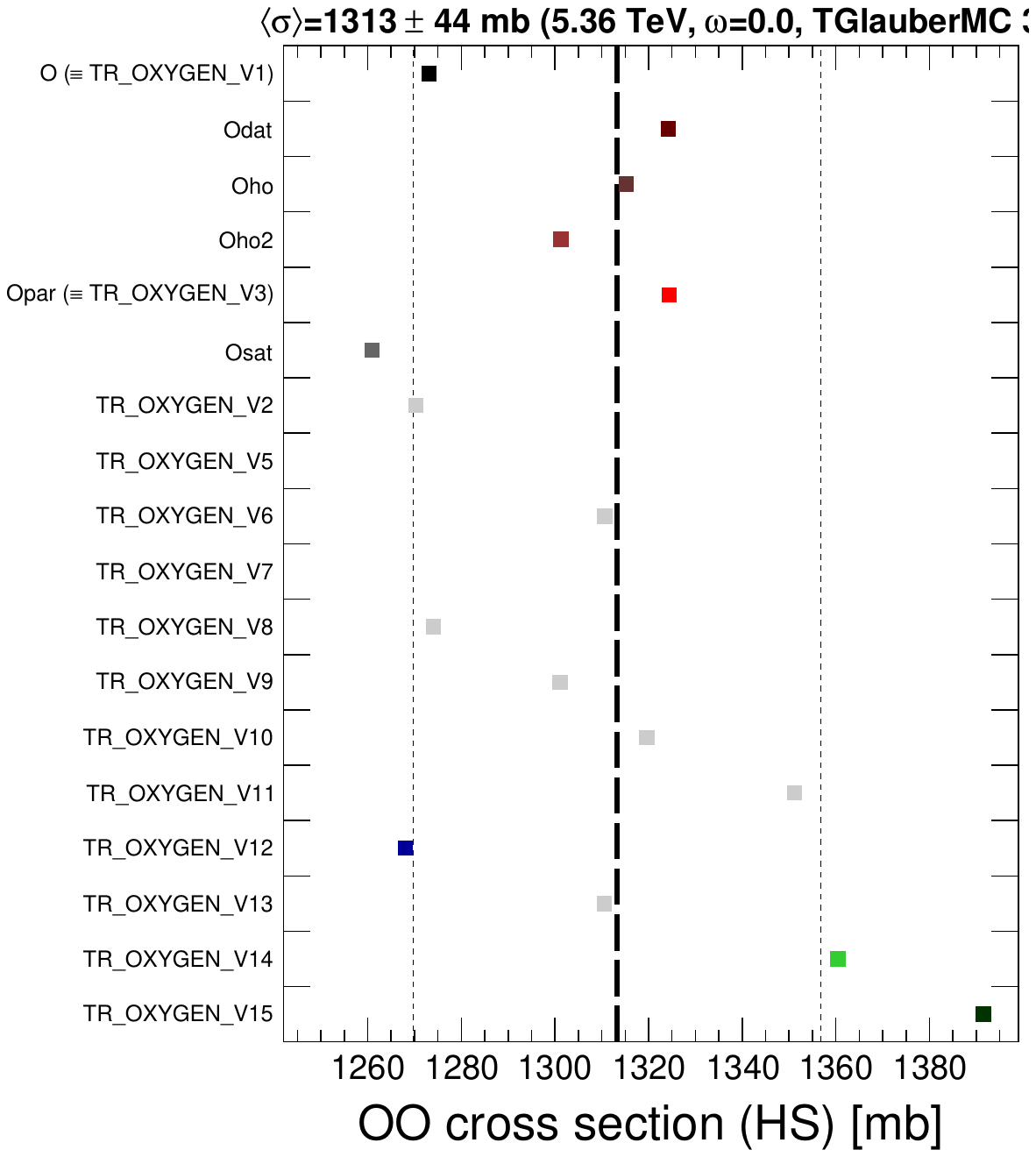}
\includegraphics[width=8cm]{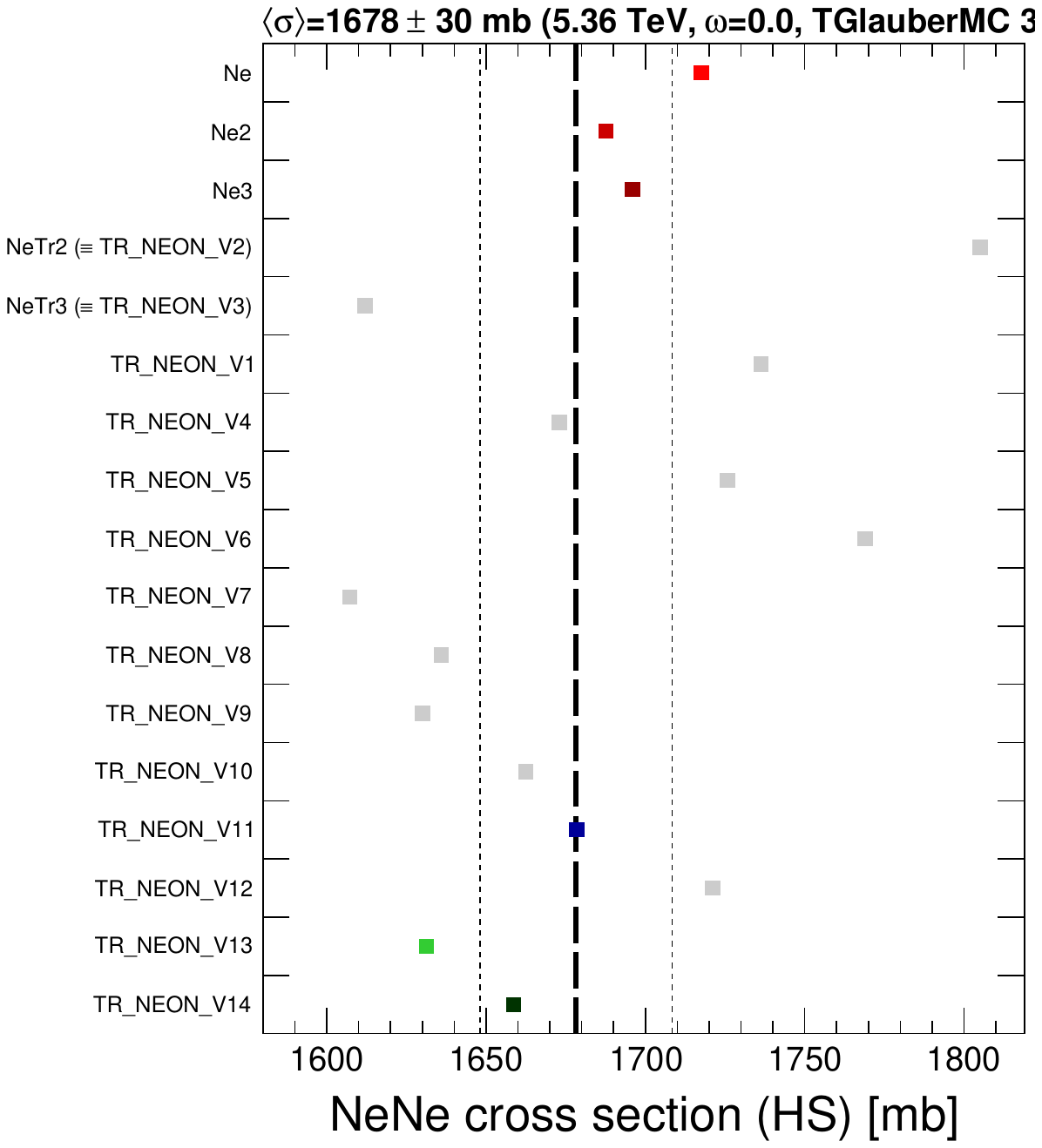}
\caption{Calculated cross sections for \OO\ and \NeNe\ collisions at 5.36~TeV using TGlauberMC with $\dmin=0.4$~fm and $\omega=0$~(hard-sphere) approximation.
The results shown in gray were obtained with non-meaningful profiles, and are not included in the determination of the average and standard deviation.}
\label{fig:extraxsec}
\end{figure}

\begin{figure}[thb!]
\includegraphics[width=8cm]{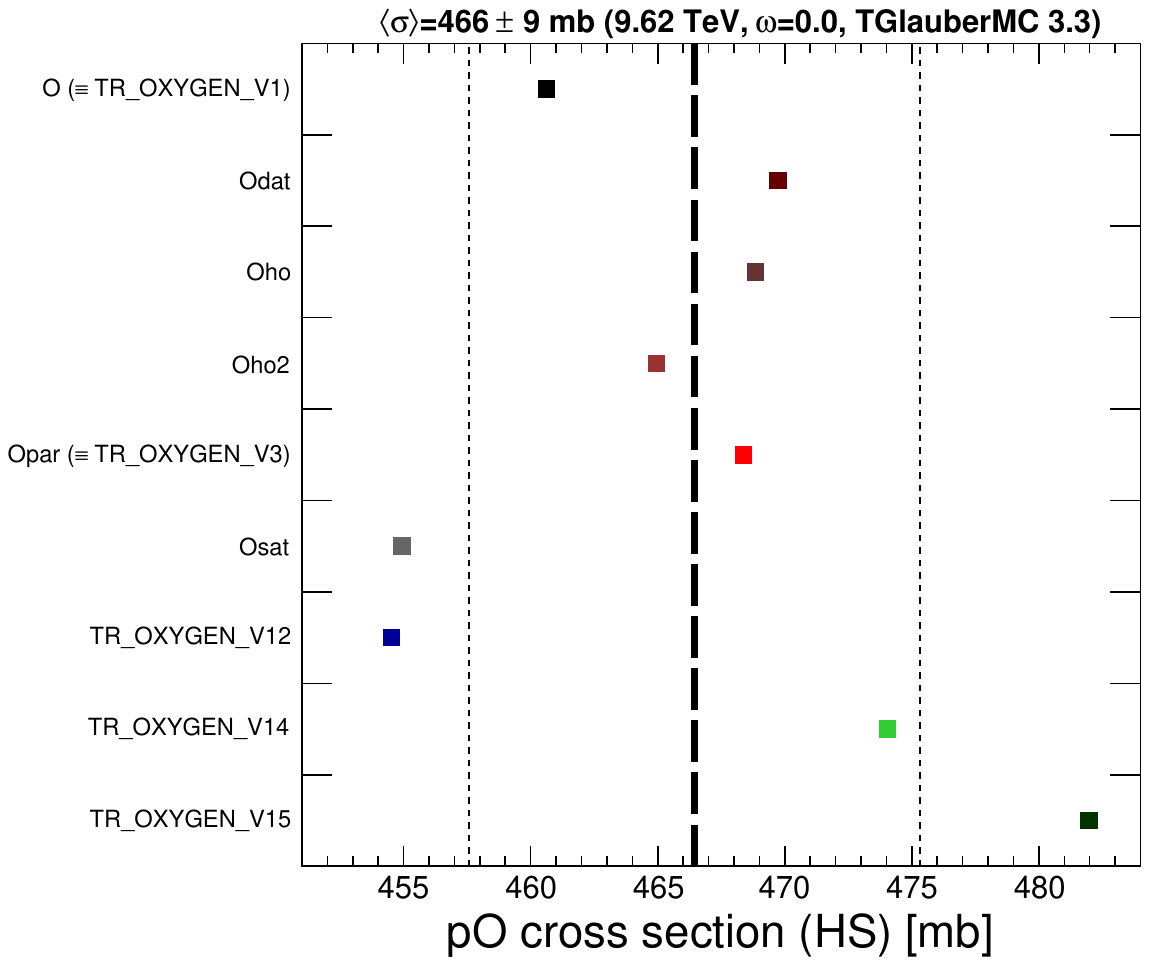}
\includegraphics[width=8cm]{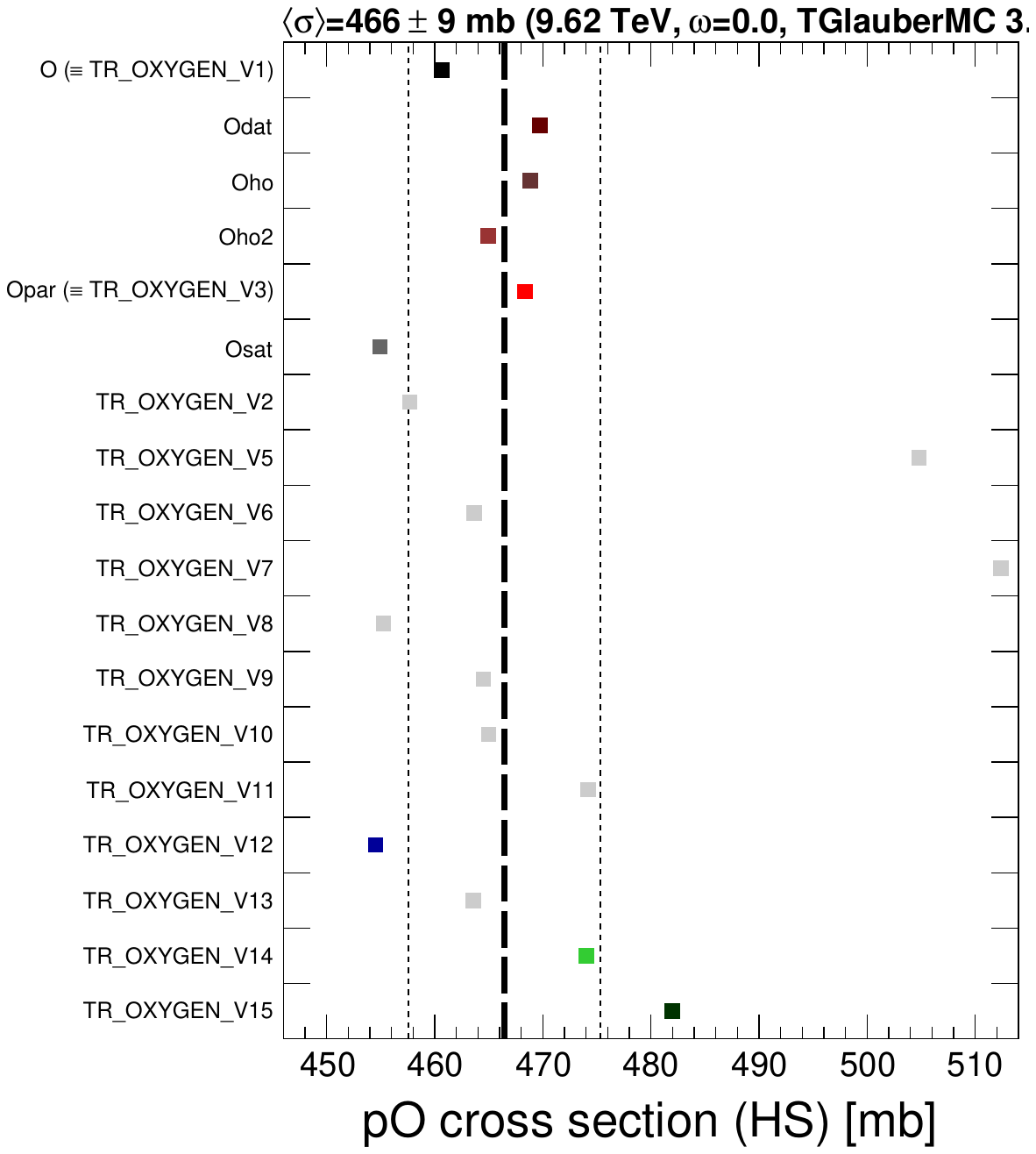}
\caption{Calculated cross sections for \pO\ collisions at 9.62 TeV using TGlauberMC with $\dmin=0.4$~fm and $\omega=0$~(hard-sphere) approximation.
The right panel shows in gray also additional results obtained with non-meaningful profiles for oxygen, and are not included in the determination of the average and standard deviation.}
\label{fig:extraxsecpo}
\end{figure}
\fi
\end{document}